\documentclass[aps,prl,reprint,preprintnumbers,superscriptaddress,amsmath,amssymb,bibnotes,longbibliography]{revtex4-2}

\usepackage{graphicx}
\usepackage{bm}
\usepackage[colorlinks,linkcolor=blue,anchorcolor=blue,citecolor=blue,urlcolor=blue,filecolor=blue,menucolor=blue,runcolor=blue]{hyperref}

\graphicspath{{figures/}}

\begin{document}


\title{Interplay of magnetism and dimerization in the pressurized Kitaev material $\beta$-Li$_2$IrO$_3$}

\author{Bin Shen}
\email{bin.shen@uni-a.de}
\affiliation  {Experimental Physics VI, Center for Electronic Correlations and Magnetism, University of Augsburg, 86159 Augsburg, Germany}

\author{Anton Jesche}
\affiliation  {Experimental Physics VI, Center for Electronic Correlations and Magnetism, University of Augsburg, 86159 Augsburg, Germany}

\author{Maximilian L. Seidler}
\affiliation  {Experimental Physics VI, Center for Electronic Correlations and Magnetism, University of Augsburg, 86159 Augsburg, Germany}

\author{Friedrich~Freund}
\affiliation  {Experimental Physics VI, Center for Electronic Correlations and Magnetism, University of Augsburg, 86159 Augsburg, Germany}

\author{Philipp Gegenwart}
\email{philipp.gegenwart@uni-a.de}
\affiliation{Experimental Physics VI, Center for Electronic Correlations and Magnetism, University of Augsburg, 86159 Augsburg, Germany}

\author{Alexander A. Tsirlin}
\email{altsirlin@gmail.com}
\affiliation{Experimental Physics VI, Center for Electronic Correlations and Magnetism, University of Augsburg, 86159 Augsburg, Germany}

\date{\today}

\begin{abstract}
We present magnetization measurements on polycrystalline $\beta$-Li$_2$IrO$_3$ under hydrostatic pressures up to 3~GPa and construct the temperature-pressure phase diagram of this material. Our data confirm that magnetic order breaks down in a first-order phase transition at $p_{\rm{c}}$ $\approx$ 1.4~GPa and additionally reveal a step-like feature -- magnetic signature of structural dimerization -- that appears at $p_{\rm{c}}$ and shifts to higher temperatures upon further compression. 
Following the structural study by L. S. I. Veiga \textit{et al.} [Phys. Rev. B \textbf{100}, 064104 (2019)], we suggest that a partially dimerized phase with a mixture of magnetic and non-magnetic Ir$^{4+}$ sites develops above $p_{\rm{c}}$. This phase is thermodynamically stable between 1.7 and 2.7~GPa according to our \textit{ab initio} calculations. It confines the magnetic Ir$^{4+}$ sites to weakly coupled tetramers with a singlet ground state and no long-range magnetic order. Our results rule out the formation of a pressure-induced spin-liquid phase in $\beta$-Li$_2$IrO$_3$ and reveal peculiarities of the magnetism collapse transition in a Kitaev material. We also show that a compressive strain imposed by the pressure treatment of $\beta$-Li$_2$IrO$_3$ enhances signatures of the 100~K magnetic anomaly at ambient pressure.
\end{abstract}

\maketitle


\section{Introduction}

The Kitaev model is arguably one of the best settings for the experimental realization of quantum spin liquid in solid-state materials~\cite{hermanns2018,19JackeliNRP}. Recent efforts have resulted in the identification of several compounds with predominant Kitaev exchange~\cite{2009JackeliPRL}, but all these compounds simultaneously showed long-range magnetic order caused by residual non-Kitaev interactions~\cite{17WinterJPCM}. External pressure was considered as a convenient tuning parameter that may reduce unwanted interactions, suppress magnetic order, and bring a material closer to the Kitaev limit~\cite{kim2016,yadav2018}. However, hydrostatic pressure experiments performed on several model compounds -- different polymorphs of Li$_2$IrO$_3$~\cite{tsirlin2021} and $\alpha$-RuCl$_3$~\cite{18BastienPRB,18BiesnerPRB,cui2017,li2019} -- all revealed a competing structural instability (dimerization) that shortens one third of the metal-metal distances on the (hyper)honeycomb spin lattice and eliminates local magnetism of the $4d/5d$ ions~\mbox{\cite{18HermannPRB,clancy2018,19TakayamaPRB,19HermannPRB}}. 

Among the model compounds tested under pressure, only $\beta$-Li$_2$IrO$_3$ with the hyperhoneycomb lattice of the Ir$^{4+}$ ions and the N\'eel temperature of $T_{\rm{N}}$ = 38~K at ambient pressure~\cite{14BiffinPRB,15TakayamaPRL,17RuizNC,19MajumderPRM}, showed some promise of entering the spin-liquid state before structural dimerization occurred. Magnetization and muon spin relaxation ($\mu$SR) measurements revealed an abrupt suppression of magnetic order at $p_{\rm{c}}$ $\approx$ 1.4~GPa, with a mixture of dynamic and randomly frozen spins above this pressure~\cite{18MajumderPRL}. In contrast, the structural dimerization was observed only at $p_{\rm dim}\simeq 4$\,GPa at room temperature~\cite{18MajumderPRL,19TakayamaPRB,choi2020}. Taken together, these results were interpreted as the gradual tuning of magnetic interactions in $\beta$-Li$_2$IrO$_3$ toward a spin-liquid phase that sets in above $p_{\rm c}$ well before the structural transformation at $p_{\rm dim}$. However, this putative spin-liquid phase must be rather fragile, as some of its spins become frozen below 15-20~K, presumably into a glassy state.

 \begin{figure}
	\includegraphics[angle=0,width=0.48\textwidth]{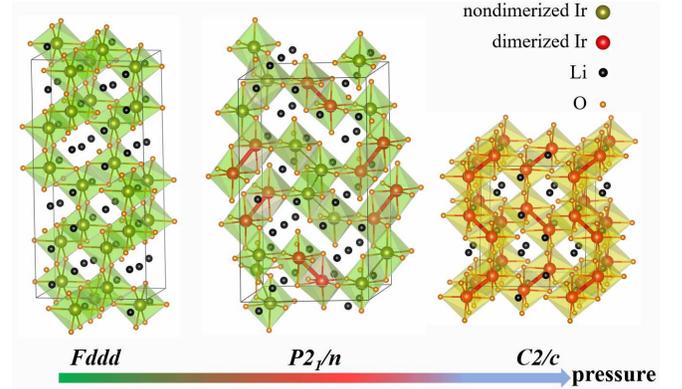}
	\vspace{-12pt} \caption{\label{figIntro} Nondimerized ($Fddd$), partially dimerized ($P2_1/n$), and fully dimerized ($C2/c$) phases of $\beta$-Li$_2$IrO$_3$ under pressure, with the crystal structures taken from Ref.~\cite{19VeigaPRB}. Red Ir--Ir bonds denote the dimerized Ir$^{4+}$ ions. }
	\vspace{-12pt}
\end{figure}

Subsequent low-temperature x-ray diffraction (XRD) experiments~\cite{19VeigaPRB} challenged this scenario. The critical pressure $p_{\rm dim}$ was shown to decrease upon cooling, and below 50~K a structural transformation was observed already around $p_{\rm c}$ concomitant with the suppression of magnetic order. However, the transformation to the dimerized phase was not completed up to at least 2.0-2.5~GPa. At low temperatures, the phase composition right above $p_{\rm c}$ is either a mixture of the fully dimerized ($C2/c$) and nondimerized ($Fddd$) phases or a distinct partially dimerized  ($P2_1/n$) phase~\cite{19VeigaPRB} (see Fig.~\ref{figIntro}). Interestingly, the $\mu$SR experiments above $p_{\rm c}$ showed neither a dimerized state nor a pure spin-liquid state but, rather, a combination of frozen and dynamic spins~\cite{18MajumderPRL} that could also be a result of multiple structural phases being present in the sample around this pressure. Moreover, no \textit{magnetic} signatures of structural dimerization in $\beta$-Li$_2$IrO$_3$ have been reported until now.

Here, we shed light on some of these peculiarities using improved magnetization measurements under pressure. We show that a step-like feature in temperature-dependent magnetic susceptibility -- the magnetic signature of structural dimerization -- appears right above $p_c$. This feature confirms that the structural transformation not only accompanies, but also triggers the suppression of the long-range magnetic order in $\beta$-Li$_2$IrO$_3$. Our data exclude the presence of the nondimerized phase above $p_{\rm c}$ and corroborate the formation of the partially dimerized phase as the most plausible state at intermediate pressures. Using \textit{ab initio} calculations we show that this phase is thermodynamically stable in a finite pressure range above $p_{\rm c}$ and features magnetic as well as non-magnetic Ir$^{4+}$ sites. The magnetic sites form weakly coupled tetramers, which are expected to show cluster magnetism and naturally evade long-range magnetic order. Intriguingly, the low-temperature susceptibility of this partially dimerized phase also shows a Curie-like upturn with the paramagnetic effective moment of about 0.7\,$\mu_B$ that persists far above $p_{\rm c}$.
 
 \begin{figure*}
 	\includegraphics[angle=0,width=0.95\textwidth]{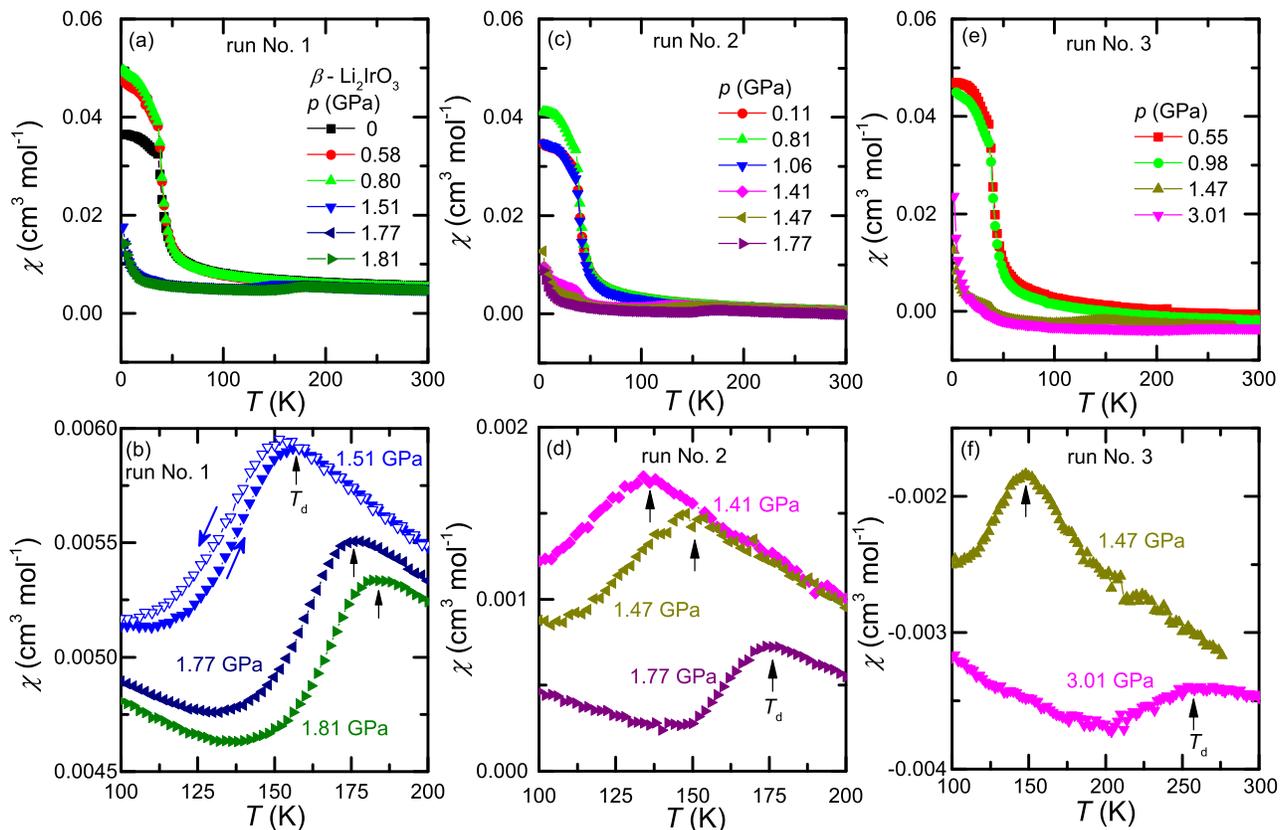}
 	\vspace{-12pt} \caption{\label{Figure1} Temperature-dependent magnetic susceptibility $\chi(T)$ of $\beta$-Li$_2$IrO$_3$ measured under various pressures from 2 to 300~K in the 1~T magnetic field. (a, c, e) Data from three different runs. (b, d, f) Magnifications of the step-like features due to the dimerization transition at $T_d$ shown by the black arrows. A hysteresis loop is clearly seen at 1.51~GPa in (b) upon cooling and warming. }
 	\vspace{-12pt}
 \end{figure*}
          
         
\section{Methods}

The polycrystalline sample of $\beta$-Li$_2$IrO$_3$ was prepared by a solid-state reaction, as described in Ref.~\cite{18MajumderPRL}. Sample quality was controlled by XDR (Rigaku MiniFlex, CuK$_{\alpha}$ radiation). Magnetization was measured using the MPMS3 SQUID magnetometer from Quantum Design. The powder sample was loaded into an opposed-anvil-type CuBe pressure cell. Measurement run Nos. 1 and 2 were carried out with a 1.8-mm anvil culet and the gasket with the sample space diameter of 0.9~mm. In this case, pressures up to 1.8~GPa could be reached. Higher pressures up to 3.0~GPa were achieved in run No.~3 with a 1-mm anvil culet and a gasket with the sample space diameter of 0.5~mm. Daphne oil 7373 was used as the pressure-transmitting medium. Pressure was determined by measuring the superconducting transition of a small piece of Pb. Magnetization of the empty cell was taken as the background. 

Pressure was applied at room temperature. The cell was cooled down to 2~K, and the data were collected upon warming unless specified otherwise. Then the pressure was increased at room temperature, and the procedure was repeated until the highest pressure feasible with the current gasket was reached. Data on decompressed samples were also collected upon warming.  

While this experimental procedure is very similar to the one implemented in Ref.~\cite{18MajumderPRL}, and the same sample and pressure cell have been used for the measurement, we took special care with the background subtraction in order to measure the weak signal above $p_{\rm c}$ with a much higher accuracy than in our previous study. We developed a software that allows the point-by-point subtraction of the background signal and an easy visual control of this procedure~\cite{Max}. Moreover, the number of raw data points for each temperature point has increased to 400 with MPMS3, compared to 48 with the much older MPMS 5\,T instrument used in Ref.~\cite{18MajumderPRL}.

High-resolution XRD data~\cite{esrf} were collected on a pristine powder sample of $\beta$-Li$_2$IrO$_3$ and on the samples recovered after decompression from run Nos. 1 and 2. XRD measurements were performed at room temperature at the ID22 beamline of the European Synchrotron Radiation Facility (ESRF; Grenoble) using the wavelength of 0.3542\,\r A and 13 scintillation detectors, each preceded by a Si (111) analyzer crystal. The samples were placed into thin-walled glass capillaries and spun during the measurement. Rietveld refinements were performed using the JANA2006 software~\cite{jana2006}.

Full-relativistic density-functional-theory (DFT) band-structure calculations were used to assess the thermodynamic stability of pressure-induced phases and their magnetism. Structural relaxations were performed in \texttt{VASP}~\cite{vasp1,vasp2} with the Perdew-Burke-Ernzerhof (PBE) for solids  (PBEsol) exchange-correlation potential~\cite{pbesol}, which yields the best agreement with the equilibrium unit cell volume of $\beta$-Li$_2$IrO$_3$ at ambient pressure. Correlation effects were taken into account on the DFT + $U$ + SO level with the on-site Coulomb repulsion parameter $U_d=1.0$\,eV and Hund's coupling $J_d=0.3$\,eV. Additionally, \texttt{FPLO} calculations~\cite{fplo} were performed on the PBE level to obtain the density of states, as well as tight-binding parameters via Wannier projections.


\section{Experimental results}
\subsection{Pressure evolution}

Figure~\ref{Figure1} shows the magnetic susceptibility $\chi$ as a function of the temperature measured under various pressures. Below 1.4~GPa, $\chi(T)$ increases smoothly upon cooling from room temperature, followed by a sharp upturn around 50~K and an anomaly at $T_{\rm{N}}$ $\approx$ 38~K due to the magnetic ordering transition. The value of $T_{\rm{N}}$ is nearly independent of the applied pressure.From the Ehrenfest relation for the specific heat and thermal expansion one estimates the initial pressure dependence of $(dT_{\rm N}/dp)_{p\rightarrow 0}=0.7$~K/GPa~\cite{18MajumderPRL}, so we expect only a marginal increase in $T_{\rm N}$ upon compression.

In the narrow pressure range between 1.4 and 1.5~GPa, the transition at $T_{\rm{N}}$ can still be  observed, but absolute values of the susceptibility become much lower. The transition coexists with a step-like feature appearing at 120-150~K. This feature is accompanied by a narrow thermal hysteresis [Fig.~\ref{Figure1}(b)] and strongly resembles the magnetic signature of structural dimerization in $\alpha$-RuCl$_3$~\cite{18BastienPRB,cui2017}. 

At even higher pressures, only the dimerization anomaly can be observed. It shifts to higher temperatures with increasing pressure [Fig.~\ref{Figure1}(b), (d), and (f)], while below this anomaly the susceptibility becomes nearly temperature-independent, except below 50~K where a Curie-like upturn appears. All these features were systematically observed in three separate measurement runs and are thus well reproducible. In  run No. 3, we used a smaller gasket and reached the pressure of 3~GPa at which the dimerization temperature $T_{\rm d}$ becomes as high as 250~K, whereas the Curie-like upturn remains nearly unchanged [Fig.~\ref{Figure1}(f)]. The slightly negative value of the susceptibility at high temperatures for run No. 3 [Fig.~\ref{Figure1}(f)] is likely due to an imperfect background subtraction caused by the more severe deformation of the gasket, since the sample volume and, thus, intrinsic magnetic signal are much reduced compared to those in run Nos. 1 and 2.

     \begin{figure}
  	\includegraphics[angle=0,width=0.49\textwidth]{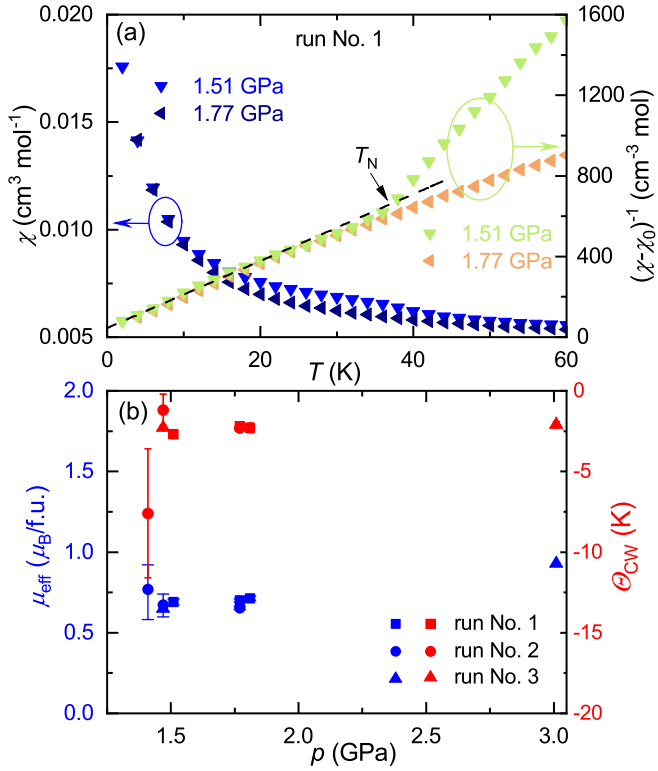}
  	\vspace{-12pt} \caption{\label{Figure2}Curie-Weiss analysis of the susceptibility data. (a) Magnetic susceptibility $\chi$ and inverse susceptibility $(\chi-\chi_0)^{-1}$ measured at 1.51 and 1.77~GPa in run No. 1. The dashed line shows the Curie-Weiss fitting at 1.51~GPa. (b) The paramagnetic effective moment and Curie-Weiss temperature $\theta_{\rm{CW}}$ extracted from the Curie-Weiss fits for the low-temperature part of the magnetic susceptibility.}
  	\vspace{-12pt}
  \end{figure}

For the Curie-Weiss analysis, we use the temperature range below 30~K [Fig.~\ref{Figure2} (a)] to avoid contributions from the paramagnetic spins of the nondimerized phase. The fit with ${\chi=\chi_0+C/(T-\theta_{\rm CW})}$ ($\chi_0$, a residual temperature-independent term; $C$, the Curie constant; $\theta_{\rm CW}$, the Curie-Weiss temperature) returns the effective moment of about 0.7~$\mu_B$/f.u. and the Curie-Weiss temperature $\theta_{\rm CW}\simeq -2$~K, both reproducible between the different measurement runs and nearly pressure independent [Fig.~\ref{Figure2}(b)]. The very small $\theta_{\rm CW}$ indicates nearly decoupled magnetic moments, which are reminiscent of impurity spins. 
For comparison, the Curie-Weiss fit of the ambient-pressure data between 200 and 300~K returns the effective moment of 1.79~$\mu_B$/f.u. in good agreement with the expected value of 1.73~$\mu_B$ for the $j_{\rm eff}=\frac12$ state of Ir$^{4+}$. Using (0.7/1.79)$^2$ = 0.152, we estimate that roughly 15~\% of the Ir atoms should be responsible for the Curie tail observed at low temperatures above $p_c$. However, this fraction may also be lower because x-ray absorption data show a rapid departure from the $j_{\rm eff}=\frac12$ state with increasing pressure~\cite{17VeigaPRB}. 

\begin{figure}
	\includegraphics[angle=0,width=0.49\textwidth]{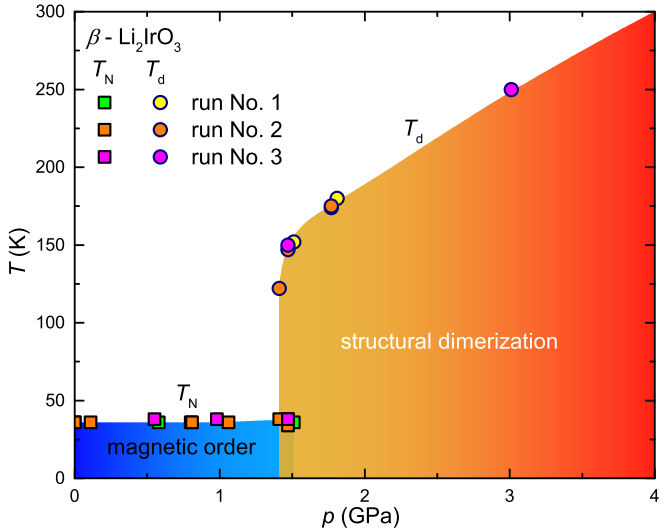}
	\vspace{-12pt} \caption{\label{Figure3} Temperature-pressure phase diagram of $\beta$-Li$_2$IrO$_3$ inferred from the magnetic susceptibility data. $T_{\rm{N}}$ marks the antiferromagnetic transition temperature of the nondimerized phase, whereas $T_{\rm{d}}$ stands for the dimerization transition temperature of the high-pressure phase. }
	\vspace{-12pt}
\end{figure}

With these revised magnetization data, we are able to confirm our earlier conclusion~\cite{18MajumderPRL} that the magnetic order in $\beta$-Li$_2$IrO$_3$ is suppressed around 1.4~GPa upon a first-order pressure-induced phase transition. However, this transition clearly has a structural component that leads to a partial dimerization with a fraction of the Ir$^{4+}$ ions paired into dimers, wherein their magnetic response is fully suppressed above $p_{\rm c}$ and below $T_{\rm d}$. Low-temperature XRD data~\cite{19VeigaPRB} suggest two possible scenarios for the phase composition above $p_{\rm c}$: (i) a mixture of the fully dimerized and nondimerized phases and (ii) a partially dimerized phase. Our results exclude the former and clearly support the latter, because the nondimerized phase with its $T_{\rm N}\simeq 38$~K should appear prominently in the magnetic susceptibility data. However, no signatures of $T_{\rm N}$ are seen above 1.51~GPa [Fig.~\ref{Figure2}(a)]. In the narrow pressure range from 1.41 to 1.51~GPa, a kink in the susceptibility is still observed because of the phase coexistence at the first-order phase transition. 

The temperature-pressure phase diagram inferred from the susceptibility data (Fig.~\ref{Figure3}) shows strong similarities to that of $\alpha$-RuCl$_3$~\cite{18BastienPRB}.The long-range-ordered phase is robust below the critical pressure $p_{\rm c}$. At $p_{\rm c}$, this phase abruptly disappears and gives way to a high-pressure phase characterized by  structural dimerization with a strong pressure dependence of $T_d$. 

Notwithstanding these apparent similarities to \mbox{$\alpha$-RuCl$_3$}, it seems pre-mature to associate the high-pressure phase of $\beta$-Li$_2$IrO$_3$ with the fully dimerized ($C2/c$) state that has been observed above 4~GPa at room temperature~\cite{18MajumderPRL,19TakayamaPRB}. 
Low-temperature XRD excludes a full transformation into the $C2/c$ phase already at $p_{\rm c}$. Moreover, the $\mu$SR observations do not support a purely non-magnetic phase above $p_{\rm c}$ and suggest a partial spin freezing below 15-20~K~\cite{18MajumderPRL}.

\begin{figure}
	\includegraphics[angle=0,width=0.49\textwidth]{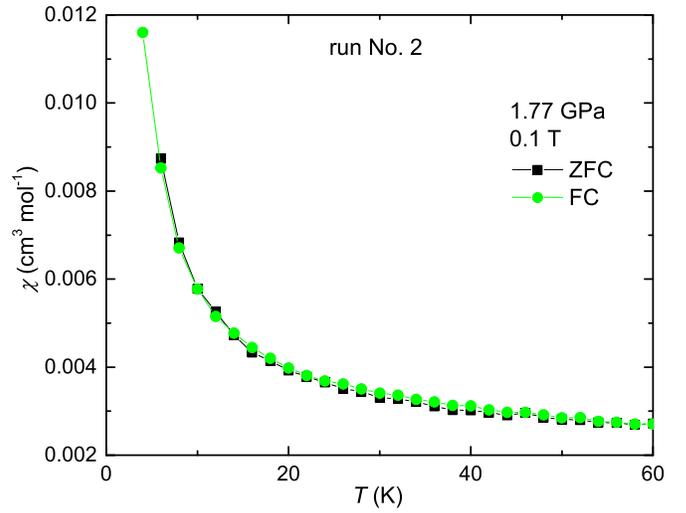}
	\vspace{-12pt} \caption{\label{Figure4}Zero-field-cooled and field-cooled magnetic susceptibility of $\beta$-Li$_2$IrO$_3$ measured at 1.77~GPa in the applied field of 0.1~T (run No. 2). No FC/ZFC splitting is observed around 15-20~K where the volume fraction of static spins increases according to $\mu$SR~\cite{18MajumderPRL}. }
\end{figure}

These $\mu$SR signatures of spin freezing should be taken with caution, though, as soon as dimerization comes into play. Freezing effects are sometimes observed in $\mu$SR experiments on dimer magnets~\cite{00AndreicaPB,03CavadiniPB,03CavadiniPB2}, presumably because muons perturb spin dimers and alter their singlet ground state. In this case, spin freezing is seen as the increase in the volume fraction of static spins in $\mu$SR but does not appear in the magnetic susceptibility. A similar situation occurs in $\beta$-Li$_2$IrO$_3$ (Fig.~\ref{Figure4}). Magnetic susceptibility measured at 1.77~GPa with $\mu_0H=0.1$~T, the lowest field that allows a reliable measurement of the weak signal in the pressure cell, does not show a splitting of the field-cooled and zero-field-cooled data around 15-20~K, the temperature range where the volume fraction of static spins increases according to $\mu$SR~\cite{18MajumderPRL}. Therefore, spin-freezing effects in the high-pressure phase of $\beta$-Li$_2$IrO$_3$ seem to be extrinsic and driven by muons.

\subsection{After decompression}

\begin{figure}
	\includegraphics[angle=0,width=0.49\textwidth]{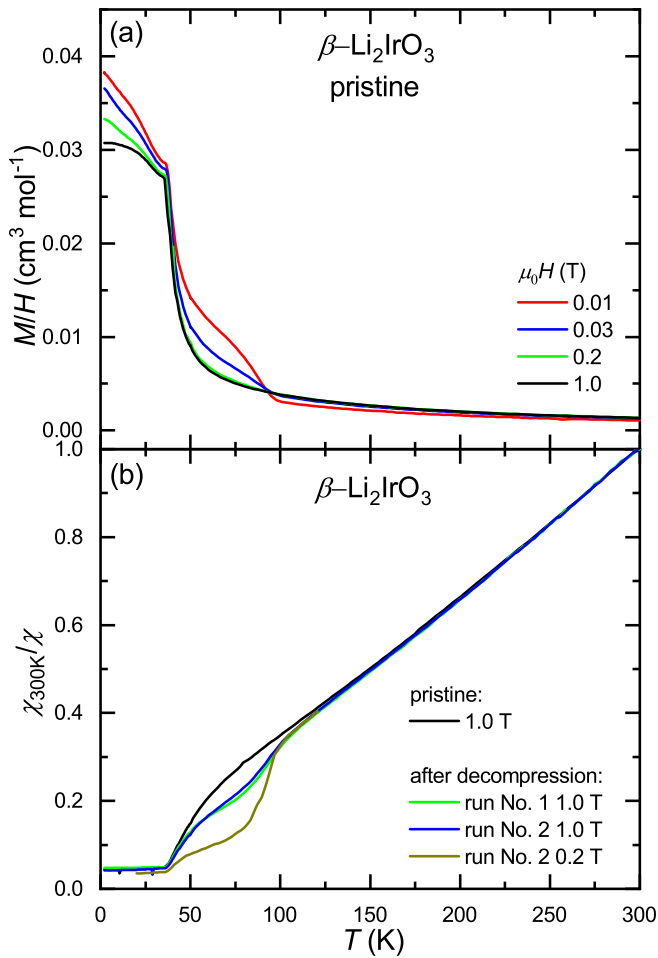}
	\vspace{-12pt} \caption{\label{Figure5}(a) Temperature-dependent magnetic susceptibility $\chi(T)$ of pristine $\beta$-Li$_2$IrO$_3$ measured upon cooling in various magnetic fields. (b) Normalized inverse magnetic susceptibility of pristine $\beta$-Li$_2$IrO$_3$, samples of run No. 1 after decompression, and samples of run No. 2 after decompression, respectively.}
\end{figure}

In Fig.~\ref{Figure5}(b), we compare the 1\,T magnetic susceptibility of the pristine $\beta$-Li$_2$IrO$_3$ sample and the samples retrieved from run NoS. 1 and 2 after decompression. Surprisingly, the anomaly at around 100\,K appears after decompression. This anomaly was also present in the pristine sample, but in much lower magnetic fields only [Fig.~\ref{Figure5}(a)]. It strongly resembles the ``high-temperature'' magnetic anomaly reported by Ruiz \textit{et al.}~\cite{ruiz2020}, who argued its intrinsic nature and a possible relation to spin-orbital magnetic species. Their data suggest that this anomaly is very sensitive to the applied magnetic field and should be fully suppressed in a field of 1~T. This is indeed the case in our pristine sample, but not in the decompressed ones, where the anomaly remains clearly visible even at 1~T.

\begin{table}
\caption{\label{tab:parameters}
Lattice parameters $a$, $b$, and $c$ and the Lorentzian profile parameter LY extracted from high-resolution powder XRD data collected on pristine and decompressed samples of $\beta$-Li$_2$IrO$_3$. Error bars are from the Rietveld refinement. The pristine sample was synthesized as powder and shows slightly different lattice parameters compared to Ref.~\cite{19MajumderPRM}, where a crushed single crystal was used.
}
\begin{ruledtabular}
\begin{tabular}{ccccc}
           & $a$ (\r A) & $b$ (\r A) & $c$ (\r A) & LY ($10^{-2}$\,deg) \\
 Pristine  & 5.91191(2) & 8.46414(2) & 17.8183(1) & 10.33(2) \\
 Run No.\,1 & 5.90737(2) & 8.45852(2) & 17.8133(1) & 11.18(2) \\
 Run No.\,2 & 5.91016(2) & 8.46208(2) & 17.8173(1) & 11.44(2) \\
\end{tabular}
\end{ruledtabular}
\end{table}

To assess structural changes introduced by the pressure treatment, we inspected the decompressed samples using high-resolution powder XRD. Rietveld refinement did not reveal any differences in the atomic positions between  decompressed and pristine samples of $\beta$-Li$_2$IrO$_3$. However, a careful comparison of the lattice and profile parameters (Table~\ref{tab:parameters}) indicated a slight increase in the reflection width gauged by the profile parameter LY and, more drastically, a systematic reduction in the lattice parameters, especially in $a$ and in $b$. This change corresponds to a small compressive strain that is more pronounced in the run No. 1 sample. Indeed, this sample showed a larger magnitude of the 100~K magnetic anomaly. While the exact origin of this anomaly remains unknown and requires a further dedicated study, we have shown that the anomaly can be enhanced by a compressive strain that, on the other hand, has no influence on $T_{\rm N}$. This indicates that long-range magnetic order in $\beta$-Li$_2$IrO$_3$ is by far more robust and less sensitive to perturbations than the elusive magnetic anomaly at 100~K.

We also note that the strain in the decompressed samples may be caused by a slight deviation from the ideal hydrostatic conditions during the pressure treatment. Nevertheless, we observe an excellent match between the magnetization data collected during run Nos. 1 and 2 despite the different amount of compressive strain accumulated in these two samples. This indicates that mo main features of the pressure-induced behavior -- the breakdown of magnetic order at $p_{\rm c}$ and the structural dimerization above $p_{\rm c}$ -- are affected by this minor deviation from the hydrostaticity.

\section{\textit{Ab initio} modeling}

\subsection{Phase stability}
We now extend this analysis with the \textit{ab initio} modeling for the partially dimerized phase ($P2_1/n$) as a plausible candidate state for $\beta$-Li$_2$IrO$_3$ above $p_{\rm c}$. The XRD data in Ref.~\cite{19VeigaPRB} reported this phase at 50~K, whereas at 25~K it could not be uniquely distinguished from a mixture of the nondimerized ($Fddd$) and fully dimerized ($C2/c$) phases. Therefore, the first question regarding the partially dimerized phase is its thermodynamic stability. Should there be a pressure range where it is more stable than both the $Fddd$ and the $C2/c$ phases, one may expect this partially dimerized phase to appear upon compression before full dimerization occurs. 

\begin{figure}
	\includegraphics{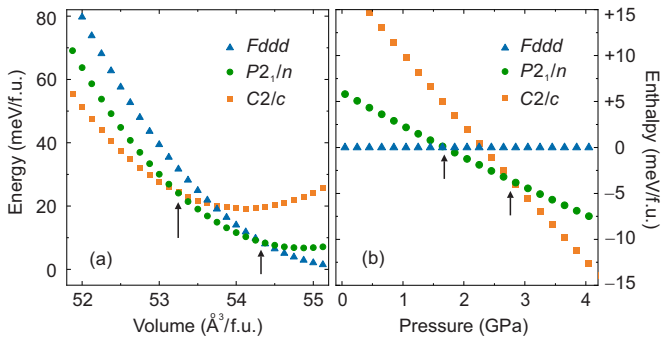}
	\caption{\label{fig:energy}
		Comparison of the nondimerized ($Fddd$), partially dimerized ($P2_1/n$), and fully dimerized ($C2/c$) phases: (a) volume dependence of energy; (b) pressure dependence of enthalpy.
	}
\end{figure}

In the following, we compare total energies and enthalpies of the phases in question. Crystal structures are relaxed at a fixed volume to obtain the volume dependence of the energy, which is fitted with the Murnaghan equation of state,
\begin{align*}
	E(V)=E_0+B_0V_0\left[\frac{1}{B_0'(B_0'-1)}\right. &\left(\frac{V}{V_0}\right)^{1-B_0'}+ \\
	&\left.+\frac{1}{B_0'}\frac{V}{V_0}-\frac{1}{B_0'-1}\right].
\end{align*}
The fit returns the equilibrium energy ($E_0$), equilibrium volume ($V_0$), and bulk modulus ($B_0$) and its pressure derivative ($B_0'$), as listed in Table~\ref{tab:eos}. 

\begin{table}
	\caption{\label{tab:eos}
		Fitted parameters of the second-order Murnaghan equation of state for different phases of $\beta$-Li$_2$IrO$_3$. Energies $E_0$ are given relative to the energy minimum of the nondimerized $Fddd$ phase.
	}
	\begin{ruledtabular}
		\begin{tabular}{ccccc}\medskip
			Space group & $E_0$ (meV/f.u.) & $V_0$ (\r A$^3$/f.u.) & $B_0$ (GPa) & $B_0'$ \\\smallskip
			$Fddd$ & 0 & 55.48(3) & 103(3) & 4.7(4) \\\smallskip
			$C2/c$ & 18(1) & 54.11(2) & 114(2) & 5.9(3) \\
			$P2_1/n$ & 6(1) & 54.91(3) & 101(3) & 6.1(3) 
		\end{tabular}
	\end{ruledtabular}
\end{table}

Figure~\ref{fig:energy} shows that the fully and partially dimerized phases are indeed stabilized under pressure. The fully dimerized phase sets in around 2.7~GPa, preceded by the partially dimerized phase, which becomes thermodynamically stable already at 1.7~GPa. These transition pressures are in very good agreement with the experimental values of, respectively, 1.5 and 2.3~GPa at 50~K, as determined from the XRD data~\cite{19VeigaPRB}. 
Our \textit{ab initio} results thus support the partially dimerized phase as a thermodynamically stable form of $\beta$-Li$_2$IrO$_3$ at intermediate pressures. It may indeed appear upon compressing $\beta$-Li$_2$IrO$_3$ at low temperatures. 

\begin{figure}
	\includegraphics{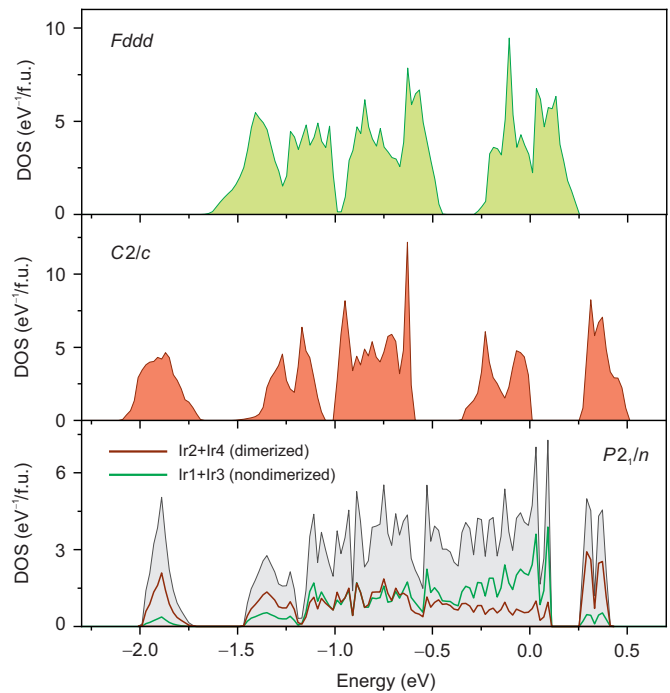}
	\caption{\label{fig:dos}
		Density of states corresponding to the Ir $t_{2g}$ bands in the three phases of $\beta$-Li$_2$IrO$_3$. Note that the partially dimerized phase ($P2_1/n$) shares features of both dimerized and nondimerized phases. The calculations are performed on the DFT + SO level without taking Coulomb correlations into account.
	}
\end{figure}

\subsection{Magnetism}
We now analyze the magnetism of this partially dimerized phase. A simple inspection of the crystal structure suggests that half of the Ir$^{4+}$ ions -- those with the Ir--Ir distance of 2.66\,\r A (Ir2 and Ir4 in the notation of Ref.~\cite{19VeigaPRB}) -- should be non-magnetic, while the remaining half (Ir1 and Ir3 with the Ir--Ir distance of 2.92\,\r A) should remain magnetic akin to the parent undimerized phase. Indeed, in DFT + $U$ + SO calculations we find magnetic moments of about 0.4\,$\mu_B$ in the nondimerized phase, no moment in the dimerized phase, and the drastically different moments of 0.05\,$\mu_B$ on Ir2 and Ir4 vs. 0.35\,$\mu_B$ on Ir1 and Ir3 in the partially dimerized phase.

This conclusion is additionally supported by the structure of the Ir $t_{2g}$ bands calculated on the DFT + SO level without taking electronic correlations into account, such that no band gap opens at the Fermi level (Fig.~\ref{fig:dos}). In the nondimerized phase, the splitting between the bands below and above $-0.5$\,eV reflects the separation of the atomic states into $j_{\rm eff}=\frac32$ and $j_{\rm eff}=\frac12$, respectively. This splitting disappears in the dimerized phase, where one finds instead several narrow bands arising from molecular orbitals of the Ir$_2$ dimers, similarly to Refs.~\cite{antonov2018} and \cite{antonov2021}. The partially dimerized phase combines both features. The Ir2 and Ir4 states participate in the dimer formation and, therefore, provide dominant contributions to the upper and lower bands around $0.3$~eV and $-1.8$~eV, respectively. The Ir1 and Ir3 states span a comparatively much narrower energy range.

\begin{figure}
	\includegraphics{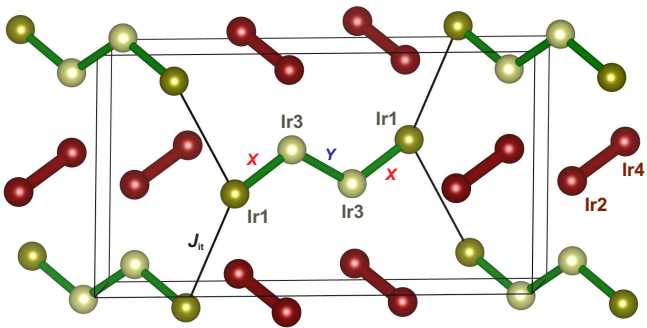}
	\caption{\label{fig:structure}
		Partially dimerized phase of $\beta$-Li$_2$IrO$_3$ with  non-magnetic Ir2--Ir4 dimers and magnetic Ir1--Ir3--Ir3--Ir1 tetramers. $J_{it}$ is the coupling between the tetramers.
	}
\end{figure}

Partial dimerization breaks the hyperhoneycomb lattice of $\beta$-Li$_2$IrO$_3$ into finite Ir1--Ir3--Ir3--Ir1 clusters with the $X-Y-X$ Kitaev bonds (Fig.~\ref{fig:structure}). The $Z$-type bonds disappear because they always involve either Ir2 or Ir4. The nature of exchange interactions is verified by a direct calculation of the exchange parameters of the spin Hamiltonian,
\begin{equation*}
	\mathcal H=\sum_{\langle ij\rangle} J_{ij}\mathbf S_i\mathbf S_j + \sum_{\langle ij\rangle} K_{ij} S_i^{\gamma}S_j^{\gamma}+\sum_{\langle ij\rangle} \Gamma_{ij} (S_i^{\alpha}S_j^{\beta}+S_i^{\beta}S_j^{\alpha}),
\end{equation*}
where $J_{ij}$, $K_{ij}$, and $\Gamma_{ij}$ stand, respectively, for the Heisenberg exchange, Kitaev exchange, and off-diagonal anisotropy, and $\alpha\neq\beta\neq\gamma$ ($\gamma=X$ for Ir1--Ir3 and $\gamma=Y$ for Ir3--Ir3). These parameters are obtained from the superexchange theory in Refs.~\cite{rau2014} and \cite{winter2016} following the procedure described in Ref.~\cite{18MajumderPRL}. 

We find $J_{13}=-6.3$\,meV, $K_{13}=-14.5$\,meV, and $\Gamma_{13}=-18.9$\,meV for the Ir1--Ir3 bonds, as well as $J_{33}=-4.4$\,meV, $K_{33}=-9.9$\,meV, and $\Gamma_{33}=-11.9$\,meV for the Ir3--Ir3 bonds. These parameters are only marginally different from the ambient-pressure values obtained using the same superexchange theory ($K=-12.1$\,meV, $\Gamma=-13.5$\,meV, and $J=-4.8$\,meV~\cite{18MajumderPRL}). Experimentally, one finds $|K|\simeq |\Gamma|\simeq 13$\,meV~\cite{19MajumderPRM,20MajumderPRB} and $J\simeq 0.3$\,meV~\cite{rousochatzakis2018}. Exact diagonalization for the Ir1--Ir3--Ir3--Ir1 tetramer was performed in Mathematica using exchange parameters for the partially dimerized phase. The energy spectrum of the tetramer features a singlet ground state separated from the first excited state by $\Delta\simeq 9.1$\,meV. For comparison, we also applied experimental exchange parameters at ambient pressure to the tetramer, and arrived at a quite similar low-energy spectrum with $\Delta\simeq 8.5$\,meV.

We thus expect that the partially dimerized phase of $\beta$-Li$_2$IrO$_3$ is magnetic, but no long-range order ensues because tetramers remain decoupled. Their singlet state is protected by the sizable gap $\Delta$. This result is consistent with our experimental observation that magnetic order vanishes above $p_{\rm c}$, but signatures of local magnetism remain visible also at higher pressures. 

The couplings between the tetramers are mediated by the non-magnetic Ir2--Ir4 dimers. The shortest superexchange pathway, with the Ir1--Ir1 distance of 5.10\,\r A, yields a coupling with $J_{\rm it}=-0.3$\,meV, $K_{\rm it}=-0.8$\,meV, and $\Gamma_{\rm it}=0.1$\,meV (Fig.~\ref{fig:structure}). This weak coupling is by far insufficient to close the gap $\Delta$ and induce long-range order.

\section{Discussion and Summary}
Our data call into question the earlier scenario of  pressure-induced spin-liquid formation in $\beta$-Li$_2$IrO$_3$~\cite{18MajumderPRL}. We have shown that the breakdown of magnetic order at $p_{\rm c}$ leads to a step-like feature in the magnetic susceptibility: a hallmark of structural dimerization. A small fraction of the nondimerized phase persists up to 1.5~GPa due to the pressure hysteresis of the first-order phase transition, but at higher pressures the magnetically ordered nondimerized phase vanishes entirely.

This leaves two possibilities for the low-temperature behavior of $\beta$-Li$_2$IrO$_3$ at pressures above $p_{\rm c}$. One scenario is the fully dimerized state, similar to the high-pressure phase of $\alpha$-RuCl$_3$, but with a significant number of defects that should account for the Curie-like upturn in the magnetic susceptibility at low temperatures. However, such a fully dimerized phase is at odds with the low-temperature XRD data at pressures right above $p_{\rm c}$~\cite{19VeigaPRB} and also fails to explain the $\mu$SR observation of mixed frozen and dynamic spins in the high-pressure phase. An interpretation of the $\mu$SR data would require that some of the muons break the singlet state of the dimers, thus causing spin freezing, while other muons leave dimers intact and observe dynamic spins. Importantly, the magnetic field distribution produced by these dynamic spins is temperature dependent and becomes broader below 40~K~\cite{18MajumderPRL}. This fact would be especially difficult to reconcile with the scenario of complete dimerization.

The alternative scenario of the partially dimerized phase seems more promising. An important aspect of this phase is that its magnetic Ir$^{4+}$ sites are confined to weakly coupled tetramer units and, therefore, evade magnetic ordering. The spins of the tetramer remain dynamic down to zero temperature, but they form a cluster magnet rather than a genuine spin liquid. The separation of the Ir$^{4+}$ ions into magnetic and non-magnetic \textit{within} one phase gives a natural explanation for the mixed $\mu$SR response of the dynamic and frozen spins coexisting in the sample. The frozen spins can be assigned to the non-magnetic Ir$^{4+}$ ions, with corresponding spin dimers perturbed by muons. The dynamic spins can be associated with the tetramers, and the temperature-dependent field distribution probed by muons may be caused by the development of spin-spin correlations on the tetramers. The 40:60 ratio of the static and dynamic spins in $\mu$SR seems rather close to the 50:50 ratio of the magnetic and non-magnetic Ir$^{4+}$ sites in the crystal structure, although some misbalance in the muon stop sites should probably be taken into account. 

One aspect of our data that even the partially dimerized phase fails to explain is the Curie-like susceptibility upturn, which is not expected in the tetramers because their ground state is a singlet. It is then natural to ascribe this Curie-like upturn to impurity spins, even if no such upturn is observed at ambient pressure or at any pressure below $p_{\rm c}$. Indeed, at ambient pressure the rapidly increasing magnetic susceptibility below $T_{\rm N}$ can mask the impurity signal or, alternatively, the impurities may become quenched when internal fields develop in the long-range-ordered state of $\beta$-Li$_2$IrO$_3$. It is also noteworthy that additional defects may be introduced by pressure treatment as compressive strain accumulates in the sample. Strain effects on the magnetism of $\beta$-Li$_2$IrO$_3$ are clearly an interesting venue for future studies. Intriguingly, the enigmatic 100~K magnetic anomaly appears to be sensitive to the strain.


Yet another aspect that requires further investigation is the transformation of the partially dimerized phase into the fully dimerized one. Our susceptibility data do not show any additional phase boundaries in the temperature-pressure phase diagram, but technical limitations restrict the highest pressure of our present study to 3~GPa, and already above 2~GPa the sensitivity of the measurement is reduced because of the smaller sample size. Magnetization measurements at higher pressures, as well as local probes, would certainly be useful to track further evolution of the unusual high-pressure phase of $\beta$-Li$_2$IrO$_3$.

In summary, we have revised the temperature-pressure phase diagram of $\beta$-Li$_2$IrO$_3$ using improved magnetization measurements. The breakdown of the long-range magnetic order around $p_{\rm c}\simeq 1.4$~GPa is accompanied by the appearance of a step-like anomaly due to the structural dimerization. This observation rules out the scenario of pressure-induced spin liquid in this material and suggests a structural instability as the main cause for the breakdown of magnetic order. The high-pressure phase above $p_{\rm c}$ can be understood as a partially dimerized phase with coexisting magnetic (nondimerized) and non-magnetic (dimerized) Ir$^{4+}$ sites.

\section{Acknowledgments}
A.A.T thanks Ioannis Rousochatzakis for his help with exact diagonalization, continuous support, and fruitful discussions. This work was funded by the German Research Foundation (DFG) via the Project No. 107745057 (TRR80) and via the Sino-German Cooperation Group on Emergent Correlated Matter.

\section{Appendix: Sample characterization}

\begin{figure}
\includegraphics{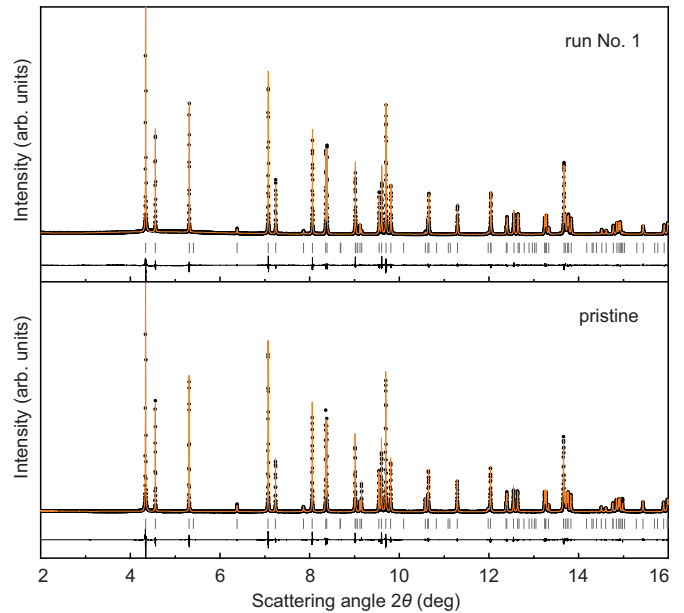}
\caption{\label{fig:refinement}
Rietveld refinements versus high-resolution XRD data. Tick marks show reflection positions for the $Fddd$ structure of nondimerized $\beta$-Li$_2$IrO$_3$. The refinement residuals are $R_I=0.028$ and $R_p=0.067$ for the sample from run No. 1 and $R_I=0.018$ and $R_p=0.080$ for the pristine sample.
}
\end{figure}

In Fig.~\ref{fig:refinement}, we show Rietveld refinements for the pristine sample and the sample recovered after decompression from run No. 1. Both samples are well described by the $Fddd$ space group of $\beta$-Li$_2$IrO$_3$, albeit with a small difference in their lattice and profile parameters (Table~\ref{tab:parameters}).

\begin{thebibliography}{40}%
	\makeatletter
	\providecommand \@ifxundefined [1]{%
		\@ifx{#1\undefined}
	}%
	\providecommand \@ifnum [1]{%
		\ifnum #1\expandafter \@firstoftwo
		\else \expandafter \@secondoftwo
		\fi
	}%
	\providecommand \@ifx [1]{%
		\ifx #1\expandafter \@firstoftwo
		\else \expandafter \@secondoftwo
		\fi
	}%
	\providecommand \natexlab [1]{#1}%
	\providecommand \enquote  [1]{``#1''}%
	\providecommand \bibnamefont  [1]{#1}%
	\providecommand \bibfnamefont [1]{#1}%
	\providecommand \citenamefont [1]{#1}%
	\providecommand \href@noop [0]{\@secondoftwo}%
	\providecommand \href [0]{\begingroup \@sanitize@url \@href}%
	\providecommand \@href[1]{\@@startlink{#1}\@@href}%
	\providecommand \@@href[1]{\endgroup#1\@@endlink}%
	\providecommand \@sanitize@url [0]{\catcode `\\12\catcode `\$12\catcode
		`\&12\catcode `\#12\catcode `\^12\catcode `\_12\catcode `\%12\relax}%
	\providecommand \@@startlink[1]{}%
	\providecommand \@@endlink[0]{}%
	\providecommand \url  [0]{\begingroup\@sanitize@url \@url }%
	\providecommand \@url [1]{\endgroup\@href {#1}{\urlprefix }}%
	\providecommand \urlprefix  [0]{URL }%
	\providecommand \Eprint [0]{\href }%
	\providecommand \doibase [0]{https://doi.org/}%
	\providecommand \selectlanguage [0]{\@gobble}%
	\providecommand \bibinfo  [0]{\@secondoftwo}%
	\providecommand \bibfield  [0]{\@secondoftwo}%
	\providecommand \translation [1]{[#1]}%
	\providecommand \BibitemOpen [0]{}%
	\providecommand \bibitemStop [0]{}%
	\providecommand \bibitemNoStop [0]{.\EOS\space}%
	\providecommand \EOS [0]{\spacefactor3000\relax}%
	\providecommand \BibitemShut  [1]{\csname bibitem#1\endcsname}%
	\let\auto@bib@innerbib\@empty
	\bibitem [{\citenamefont {Hermanns}\ \emph {et~al.}(2018)\citenamefont
		{Hermanns}, \citenamefont {Kimchi},\ and\ \citenamefont
		{Knolle}}]{hermanns2018}%
	\BibitemOpen
	\bibfield  {author} {\bibinfo {author} {\bibfnamefont {M.}~\bibnamefont
			{Hermanns}}, \bibinfo {author} {\bibfnamefont {I.}~\bibnamefont {Kimchi}},\
		and\ \bibinfo {author} {\bibfnamefont {J.}~\bibnamefont {Knolle}},\
	}\bibfield  {title} {\bibinfo {title} {Physics of the {Kitaev} model:
			Fractionalization, dynamic correlations, and material connections},\ }\href
	{https://doi.org/10.1146/annurev-conmatphys-033117-053934} {\bibfield
		{journal} {\bibinfo  {journal} {Ann. Rev. Condens. Matter Phys.}\ }\textbf
		{\bibinfo {volume} {9}},\ \bibinfo {pages} {17} (\bibinfo {year}
		{2018})}\BibitemShut {NoStop}%
	\bibitem [{\citenamefont {Takagi}\ \emph {et~al.}(2019)\citenamefont {Takagi},
		\citenamefont {Takayama}, \citenamefont {Jackeli}, \citenamefont
		{Khaliullin},\ and\ \citenamefont {Nagler}}]{19JackeliNRP}%
	\BibitemOpen
	\bibfield  {author} {\bibinfo {author} {\bibfnamefont {H.}~\bibnamefont
			{Takagi}}, \bibinfo {author} {\bibfnamefont {T.}~\bibnamefont {Takayama}},
		\bibinfo {author} {\bibfnamefont {G.}~\bibnamefont {Jackeli}}, \bibinfo
		{author} {\bibfnamefont {G.}~\bibnamefont {Khaliullin}},\ and\ \bibinfo
		{author} {\bibfnamefont {S.~E.}\ \bibnamefont {Nagler}},\ }\bibfield  {title}
	{\bibinfo {title} {Concept and realization of {Kitaev} quantum spin
			liquids},\ }\href {https://doi.org/10.1038/s42254-019-0038-2} {\bibfield
		{journal} {\bibinfo  {journal} {Nature Reviews Physics}\ }\textbf {\bibinfo
			{volume} {1}},\ \bibinfo {pages} {264} (\bibinfo {year} {2019})}\BibitemShut
	{NoStop}%
	\bibitem [{\citenamefont {Jackeli}\ and\ \citenamefont
		{Khaliullin}(2009)}]{2009JackeliPRL}%
	\BibitemOpen
	\bibfield  {author} {\bibinfo {author} {\bibfnamefont {G.}~\bibnamefont
			{Jackeli}}\ and\ \bibinfo {author} {\bibfnamefont {G.}~\bibnamefont
			{Khaliullin}},\ }\bibfield  {title} {\bibinfo {title} {Mott insulators in the
			strong spin-orbit coupling limit: From {Heisenberg} to a quantum compass and
			{Kitaev} models},\ }\href {https://doi.org/10.1103/PhysRevLett.102.017205}
	{\bibfield  {journal} {\bibinfo  {journal} {Phys. Rev. Lett.}\ }\textbf
		{\bibinfo {volume} {102}},\ \bibinfo {pages} {017205} (\bibinfo {year}
		{2009})}\BibitemShut {NoStop}%
	\bibitem [{\citenamefont {Winter}\ \emph {et~al.}(2017)\citenamefont {Winter},
		\citenamefont {Tsirlin}, \citenamefont {Daghofer}, \citenamefont {van~den
			Brink}, \citenamefont {Singh}, \citenamefont {Gegenwart},\ and\ \citenamefont
		{Valent{\'{\i}}}}]{17WinterJPCM}%
	\BibitemOpen
	\bibfield  {author} {\bibinfo {author} {\bibfnamefont {S.~M.}\ \bibnamefont
			{Winter}}, \bibinfo {author} {\bibfnamefont {A.~A.}\ \bibnamefont {Tsirlin}},
		\bibinfo {author} {\bibfnamefont {M.}~\bibnamefont {Daghofer}}, \bibinfo
		{author} {\bibfnamefont {J.}~\bibnamefont {van~den Brink}}, \bibinfo {author}
		{\bibfnamefont {Y.}~\bibnamefont {Singh}}, \bibinfo {author} {\bibfnamefont
			{P.}~\bibnamefont {Gegenwart}},\ and\ \bibinfo {author} {\bibfnamefont
			{R.}~\bibnamefont {Valent{\'{\i}}}},\ }\bibfield  {title} {\bibinfo {title}
		{Models and materials for generalized {Kitaev} magnetism},\ }\href
	{https://doi.org/10.1088/1361-648x/aa8cf5} {\bibfield  {journal} {\bibinfo
			{journal} {J. Phys.: Condens. Matter}\ }\textbf {\bibinfo {volume} {29}},\
		\bibinfo {pages} {493002} (\bibinfo {year} {2017})}\BibitemShut {NoStop}%
	\bibitem [{\citenamefont {Kim}\ \emph {et~al.}(2016)\citenamefont {Kim},
		\citenamefont {Kim},\ and\ \citenamefont {Kee}}]{kim2016}%
	\BibitemOpen
	\bibfield  {author} {\bibinfo {author} {\bibfnamefont {H.-S.}\ \bibnamefont
			{Kim}}, \bibinfo {author} {\bibfnamefont {Y.~B.}\ \bibnamefont {Kim}},\ and\
		\bibinfo {author} {\bibfnamefont {H.-Y.}\ \bibnamefont {Kee}},\ }\bibfield
	{title} {\bibinfo {title} {Revealing frustrated local moment model for
			pressurized hyperhoneycomb iridate: Paving the way toward a quantum spin
			liquid},\ }\href {https://doi.org/10.1103/PhysRevB.94.245127} {\bibfield
		{journal} {\bibinfo  {journal} {Phys. Rev. B}\ }\textbf {\bibinfo {volume}
			{94}},\ \bibinfo {pages} {245127} (\bibinfo {year} {2016})}\BibitemShut
	{NoStop}%
	\bibitem [{\citenamefont {Yadav}\ \emph {et~al.}(2018)\citenamefont {Yadav},
		\citenamefont {Rachel}, \citenamefont {Hozoi}, \citenamefont {van~den
			Brink},\ and\ \citenamefont {Jackeli}}]{yadav2018}%
	\BibitemOpen
	\bibfield  {author} {\bibinfo {author} {\bibfnamefont {R.}~\bibnamefont
			{Yadav}}, \bibinfo {author} {\bibfnamefont {S.}~\bibnamefont {Rachel}},
		\bibinfo {author} {\bibfnamefont {L.}~\bibnamefont {Hozoi}}, \bibinfo
		{author} {\bibfnamefont {J.}~\bibnamefont {van~den Brink}},\ and\ \bibinfo
		{author} {\bibfnamefont {G.}~\bibnamefont {Jackeli}},\ }\bibfield  {title}
	{\bibinfo {title} {Strain- and pressure-tuned magnetic interactions in
			honeycomb {Kitaev} materials},\ }\href
	{https://doi.org/10.1103/PhysRevB.98.121107} {\bibfield  {journal} {\bibinfo
			{journal} {Phys. Rev. B}\ }\textbf {\bibinfo {volume} {98}},\ \bibinfo
		{pages} {121107(R)} (\bibinfo {year} {2018})}\BibitemShut {NoStop}%
	\bibitem [{\citenamefont {Tsirlin}\ and\ \citenamefont
		{Gegenwart}(2021)}]{tsirlin2021}%
	\BibitemOpen
	\bibfield  {author} {\bibinfo {author} {\bibfnamefont {A.~A.}\ \bibnamefont
			{Tsirlin}}\ and\ \bibinfo {author} {\bibfnamefont {P.}~\bibnamefont
			{Gegenwart}},\ }\bibfield  {title} {\bibinfo {title} {Kitaev magnetism
			through the prism of lithium iridate},\ }\href
	{https://doi.org/10.1002/pssb.202100146} {\bibfield  {journal} {\bibinfo
			{journal} {Phys. Status Solidi B}\ ,\ \bibinfo {pages} {2100146}} (\bibinfo
		{year} {2021})}\BibitemShut {NoStop}%
	\bibitem [{\citenamefont {Bastien}\ \emph {et~al.}(2018)\citenamefont
		{Bastien}, \citenamefont {Garbarino}, \citenamefont {Yadav}, \citenamefont
		{Martinez-Casado}, \citenamefont {Beltr\'an~Rodr\'{\i}guez}, \citenamefont
		{Stahl}, \citenamefont {Kusch}, \citenamefont {Limandri}, \citenamefont
		{Ray}, \citenamefont {Lampen-Kelley}, \citenamefont {Mandrus}, \citenamefont
		{Nagler}, \citenamefont {Roslova}, \citenamefont {Isaeva}, \citenamefont
		{Doert}, \citenamefont {Hozoi}, \citenamefont {Wolter}, \citenamefont
		{B\"uchner}, \citenamefont {Geck},\ and\ \citenamefont {van~den
			Brink}}]{18BastienPRB}%
	\BibitemOpen
	\bibfield  {author} {\bibinfo {author} {\bibfnamefont {G.}~\bibnamefont
			{Bastien}}, \bibinfo {author} {\bibfnamefont {G.}~\bibnamefont {Garbarino}},
		\bibinfo {author} {\bibfnamefont {R.}~\bibnamefont {Yadav}}, \bibinfo
		{author} {\bibfnamefont {F.~J.}\ \bibnamefont {Martinez-Casado}}, \bibinfo
		{author} {\bibfnamefont {R.}~\bibnamefont {Beltr\'an~Rodr\'{\i}guez}},
		\bibinfo {author} {\bibfnamefont {Q.}~\bibnamefont {Stahl}}, \bibinfo
		{author} {\bibfnamefont {M.}~\bibnamefont {Kusch}}, \bibinfo {author}
		{\bibfnamefont {S.~P.}\ \bibnamefont {Limandri}}, \bibinfo {author}
		{\bibfnamefont {R.}~\bibnamefont {Ray}}, \bibinfo {author} {\bibfnamefont
			{P.}~\bibnamefont {Lampen-Kelley}}, \bibinfo {author} {\bibfnamefont {D.~G.}\
			\bibnamefont {Mandrus}}, \bibinfo {author} {\bibfnamefont {S.~E.}\
			\bibnamefont {Nagler}}, \bibinfo {author} {\bibfnamefont {M.}~\bibnamefont
			{Roslova}}, \bibinfo {author} {\bibfnamefont {A.}~\bibnamefont {Isaeva}},
		\bibinfo {author} {\bibfnamefont {T.}~\bibnamefont {Doert}}, \bibinfo
		{author} {\bibfnamefont {L.}~\bibnamefont {Hozoi}}, \bibinfo {author}
		{\bibfnamefont {A.~U.~B.}\ \bibnamefont {Wolter}}, \bibinfo {author}
		{\bibfnamefont {B.}~\bibnamefont {B\"uchner}}, \bibinfo {author}
		{\bibfnamefont {J.}~\bibnamefont {Geck}},\ and\ \bibinfo {author}
		{\bibfnamefont {J.}~\bibnamefont {van~den Brink}},\ }\bibfield  {title}
	{\bibinfo {title} {Pressure-induced dimerization and valence bond crystal
			formation in the {Kitaev-Heisenberg} magnet
			$\ensuremath{\alpha}\text{\ensuremath{-}}\mathrm{RuCl}{}_{3}$},\ }\href
	{https://doi.org/10.1103/PhysRevB.97.241108} {\bibfield  {journal} {\bibinfo
			{journal} {Phys. Rev. B}\ }\textbf {\bibinfo {volume} {97}},\ \bibinfo
		{pages} {241108} (\bibinfo {year} {2018})}\BibitemShut {NoStop}%
	\bibitem [{\citenamefont {Biesner}\ \emph {et~al.}(2018)\citenamefont
		{Biesner}, \citenamefont {Biswas}, \citenamefont {Li}, \citenamefont {Saito},
		\citenamefont {Pustogow}, \citenamefont {Altmeyer}, \citenamefont {Wolter},
		\citenamefont {B\"uchner}, \citenamefont {Roslova}, \citenamefont {Doert},
		\citenamefont {Winter}, \citenamefont {Valent\'{\i}},\ and\ \citenamefont
		{Dressel}}]{18BiesnerPRB}%
	\BibitemOpen
	\bibfield  {author} {\bibinfo {author} {\bibfnamefont {T.}~\bibnamefont
			{Biesner}}, \bibinfo {author} {\bibfnamefont {S.}~\bibnamefont {Biswas}},
		\bibinfo {author} {\bibfnamefont {W.}~\bibnamefont {Li}}, \bibinfo {author}
		{\bibfnamefont {Y.}~\bibnamefont {Saito}}, \bibinfo {author} {\bibfnamefont
			{A.}~\bibnamefont {Pustogow}}, \bibinfo {author} {\bibfnamefont
			{M.}~\bibnamefont {Altmeyer}}, \bibinfo {author} {\bibfnamefont {A.~U.~B.}\
			\bibnamefont {Wolter}}, \bibinfo {author} {\bibfnamefont {B.}~\bibnamefont
			{B\"uchner}}, \bibinfo {author} {\bibfnamefont {M.}~\bibnamefont {Roslova}},
		\bibinfo {author} {\bibfnamefont {T.}~\bibnamefont {Doert}}, \bibinfo
		{author} {\bibfnamefont {S.~M.}\ \bibnamefont {Winter}}, \bibinfo {author}
		{\bibfnamefont {R.}~\bibnamefont {Valent\'{\i}}},\ and\ \bibinfo {author}
		{\bibfnamefont {M.}~\bibnamefont {Dressel}},\ }\bibfield  {title} {\bibinfo
		{title} {Detuning the honeycomb of {$\alpha$-RuCl$_3$}: Pressure-dependent
			optical studies reveal broken symmetry},\ }\href
	{https://doi.org/10.1103/PhysRevB.97.220401} {\bibfield  {journal} {\bibinfo
			{journal} {Phys. Rev. B}\ }\textbf {\bibinfo {volume} {97}},\ \bibinfo
		{pages} {220401} (\bibinfo {year} {2018})}\BibitemShut {NoStop}%
	\bibitem [{\citenamefont {Cui}\ \emph {et~al.}(2017)\citenamefont {Cui},
		\citenamefont {Zheng}, \citenamefont {Ran}, \citenamefont {Wen},
		\citenamefont {Liu}, \citenamefont {Liu}, \citenamefont {Guo},\ and\
		\citenamefont {Yu}}]{cui2017}%
	\BibitemOpen
	\bibfield  {author} {\bibinfo {author} {\bibfnamefont {Y.}~\bibnamefont
			{Cui}}, \bibinfo {author} {\bibfnamefont {J.}~\bibnamefont {Zheng}}, \bibinfo
		{author} {\bibfnamefont {K.}~\bibnamefont {Ran}}, \bibinfo {author}
		{\bibfnamefont {J.}~\bibnamefont {Wen}}, \bibinfo {author} {\bibfnamefont
			{Z.-X.}\ \bibnamefont {Liu}}, \bibinfo {author} {\bibfnamefont
			{B.}~\bibnamefont {Liu}}, \bibinfo {author} {\bibfnamefont {W.}~\bibnamefont
			{Guo}},\ and\ \bibinfo {author} {\bibfnamefont {W.}~\bibnamefont {Yu}},\
	}\bibfield  {title} {\bibinfo {title} {High-pressure magnetization and {NMR}
			studies of {$\alpha$-RuCl$_3$}},\ }\href
	{https://doi.org/10.1103/PhysRevB.96.205147} {\bibfield  {journal} {\bibinfo
			{journal} {Phys. Rev. B}\ }\textbf {\bibinfo {volume} {96}},\ \bibinfo
		{pages} {205147} (\bibinfo {year} {2017})}\BibitemShut {NoStop}%
	\bibitem [{\citenamefont {Li}\ \emph {et~al.}(2019)\citenamefont {Li},
		\citenamefont {Chen}, \citenamefont {Gan}, \citenamefont {Li}, \citenamefont
		{Yan}, \citenamefont {Ye}, \citenamefont {Pei}, \citenamefont {Zhang},
		\citenamefont {Wang}, \citenamefont {Su}, \citenamefont {Dai}, \citenamefont
		{Chen}, \citenamefont {Shi}, \citenamefont {Wang}, \citenamefont {Zhang},
		\citenamefont {Wang}, \citenamefont {Yu}, \citenamefont {Ye}, \citenamefont
		{Mei},\ and\ \citenamefont {Huang}}]{li2019}%
	\BibitemOpen
	\bibfield  {author} {\bibinfo {author} {\bibfnamefont {G.}~\bibnamefont
			{Li}}, \bibinfo {author} {\bibfnamefont {X.}~\bibnamefont {Chen}}, \bibinfo
		{author} {\bibfnamefont {Y.}~\bibnamefont {Gan}}, \bibinfo {author}
		{\bibfnamefont {F.}~\bibnamefont {Li}}, \bibinfo {author} {\bibfnamefont
			{M.}~\bibnamefont {Yan}}, \bibinfo {author} {\bibfnamefont {F.}~\bibnamefont
			{Ye}}, \bibinfo {author} {\bibfnamefont {S.}~\bibnamefont {Pei}}, \bibinfo
		{author} {\bibfnamefont {Y.}~\bibnamefont {Zhang}}, \bibinfo {author}
		{\bibfnamefont {L.}~\bibnamefont {Wang}}, \bibinfo {author} {\bibfnamefont
			{H.}~\bibnamefont {Su}}, \bibinfo {author} {\bibfnamefont {J.}~\bibnamefont
			{Dai}}, \bibinfo {author} {\bibfnamefont {Y.}~\bibnamefont {Chen}}, \bibinfo
		{author} {\bibfnamefont {Y.}~\bibnamefont {Shi}}, \bibinfo {author}
		{\bibfnamefont {X.}~\bibnamefont {Wang}}, \bibinfo {author} {\bibfnamefont
			{L.}~\bibnamefont {Zhang}}, \bibinfo {author} {\bibfnamefont
			{S.}~\bibnamefont {Wang}}, \bibinfo {author} {\bibfnamefont {D.}~\bibnamefont
			{Yu}}, \bibinfo {author} {\bibfnamefont {F.}~\bibnamefont {Ye}}, \bibinfo
		{author} {\bibfnamefont {J.-W.}\ \bibnamefont {Mei}},\ and\ \bibinfo {author}
		{\bibfnamefont {M.}~\bibnamefont {Huang}},\ }\bibfield  {title} {\bibinfo
		{title} {Raman spectroscopy evidence for dimerization and {Mott} collapse in
			{$\alpha$-RuCl$_3$} under pressures},\ }\href
	{https://doi.org/10.1103/PhysRevMaterials.3.023601} {\bibfield  {journal}
		{\bibinfo  {journal} {Phys. Rev. Materials}\ }\textbf {\bibinfo {volume}
			{3}},\ \bibinfo {pages} {023601} (\bibinfo {year} {2019})}\BibitemShut
	{NoStop}%
	\bibitem [{\citenamefont {Hermann}\ \emph {et~al.}(2018)\citenamefont
		{Hermann}, \citenamefont {Altmeyer}, \citenamefont {Ebad-Allah},
		\citenamefont {Freund}, \citenamefont {Jesche}, \citenamefont {Tsirlin},
		\citenamefont {Hanfland}, \citenamefont {Gegenwart}, \citenamefont {Mazin},
		\citenamefont {Khomskii}, \citenamefont {Valent\'{\i}},\ and\ \citenamefont
		{Kuntscher}}]{18HermannPRB}%
	\BibitemOpen
	\bibfield  {author} {\bibinfo {author} {\bibfnamefont {V.}~\bibnamefont
			{Hermann}}, \bibinfo {author} {\bibfnamefont {M.}~\bibnamefont {Altmeyer}},
		\bibinfo {author} {\bibfnamefont {J.}~\bibnamefont {Ebad-Allah}}, \bibinfo
		{author} {\bibfnamefont {F.}~\bibnamefont {Freund}}, \bibinfo {author}
		{\bibfnamefont {A.}~\bibnamefont {Jesche}}, \bibinfo {author} {\bibfnamefont
			{A.~A.}\ \bibnamefont {Tsirlin}}, \bibinfo {author} {\bibfnamefont
			{M.}~\bibnamefont {Hanfland}}, \bibinfo {author} {\bibfnamefont
			{P.}~\bibnamefont {Gegenwart}}, \bibinfo {author} {\bibfnamefont {I.~I.}\
			\bibnamefont {Mazin}}, \bibinfo {author} {\bibfnamefont {D.~I.}\ \bibnamefont
			{Khomskii}}, \bibinfo {author} {\bibfnamefont {R.}~\bibnamefont
			{Valent\'{\i}}},\ and\ \bibinfo {author} {\bibfnamefont {C.~A.}\ \bibnamefont
			{Kuntscher}},\ }\bibfield  {title} {\bibinfo {title} {Competition between
			spin-orbit coupling, magnetism, and dimerization in the honeycomb iridates:
			$\ensuremath{\alpha}\text{\ensuremath{-}}\mathrm{Li}{}_{2}\mathrm{IrO}{}_{3}$
			under pressure},\ }\href {https://doi.org/10.1103/PhysRevB.97.020104}
	{\bibfield  {journal} {\bibinfo  {journal} {Phys. Rev. B}\ }\textbf {\bibinfo
			{volume} {97}},\ \bibinfo {pages} {020104} (\bibinfo {year}
		{2018})}\BibitemShut {NoStop}%
	\bibitem [{\citenamefont {Clancy}\ \emph {et~al.}(2018)\citenamefont {Clancy},
		\citenamefont {Gretarsson}, \citenamefont {Sears}, \citenamefont {Singh},
		\citenamefont {Desgreniers}, \citenamefont {Mehlawat}, \citenamefont {Layek},
		\citenamefont {Rozenberg}, \citenamefont {Ding}, \citenamefont {Upton},
		\citenamefont {Casa}, \citenamefont {Chen}, \citenamefont {Im}, \citenamefont
		{Lee}, \citenamefont {Yadav}, \citenamefont {Hozoi}, \citenamefont {Efremov},
		\citenamefont {{van den Brink}},\ and\ \citenamefont {Kim}}]{clancy2018}%
	\BibitemOpen
	\bibfield  {author} {\bibinfo {author} {\bibfnamefont {J.~P.}\ \bibnamefont
			{Clancy}}, \bibinfo {author} {\bibfnamefont {H.}~\bibnamefont {Gretarsson}},
		\bibinfo {author} {\bibfnamefont {J.~A.}\ \bibnamefont {Sears}}, \bibinfo
		{author} {\bibfnamefont {Y.}~\bibnamefont {Singh}}, \bibinfo {author}
		{\bibfnamefont {S.}~\bibnamefont {Desgreniers}}, \bibinfo {author}
		{\bibfnamefont {K.}~\bibnamefont {Mehlawat}}, \bibinfo {author}
		{\bibfnamefont {S.}~\bibnamefont {Layek}}, \bibinfo {author} {\bibfnamefont
			{G.~K.}\ \bibnamefont {Rozenberg}}, \bibinfo {author} {\bibfnamefont
			{Y.}~\bibnamefont {Ding}}, \bibinfo {author} {\bibfnamefont {M.~H.}\
			\bibnamefont {Upton}}, \bibinfo {author} {\bibfnamefont {D.}~\bibnamefont
			{Casa}}, \bibinfo {author} {\bibfnamefont {N.}~\bibnamefont {Chen}}, \bibinfo
		{author} {\bibfnamefont {J.}~\bibnamefont {Im}}, \bibinfo {author}
		{\bibfnamefont {Y.}~\bibnamefont {Lee}}, \bibinfo {author} {\bibfnamefont
			{R.}~\bibnamefont {Yadav}}, \bibinfo {author} {\bibfnamefont
			{L.}~\bibnamefont {Hozoi}}, \bibinfo {author} {\bibfnamefont
			{D.}~\bibnamefont {Efremov}}, \bibinfo {author} {\bibfnamefont
			{J.}~\bibnamefont {{van den Brink}}},\ and\ \bibinfo {author} {\bibfnamefont
			{Y.-J.}\ \bibnamefont {Kim}},\ }\bibfield  {title} {\bibinfo {title}
		{Pressure-driven collapse of the relativistic electronic ground state in a
			honeycomb iridate},\ }\href {https://doi.org/10.1038/s41535-018-0109-0}
	{\bibfield  {journal} {\bibinfo  {journal} {npj Quantum Materials}\ }\textbf
		{\bibinfo {volume} {3}},\ \bibinfo {pages} {35} (\bibinfo {year}
		{2018})}\BibitemShut {NoStop}%
	\bibitem [{\citenamefont {Takayama}\ \emph {et~al.}(2019)\citenamefont
		{Takayama}, \citenamefont {Krajewska}, \citenamefont {Gibbs}, \citenamefont
		{Yaresko}, \citenamefont {Ishii}, \citenamefont {Yamaoka}, \citenamefont
		{Ishii}, \citenamefont {Hiraoka}, \citenamefont {Funnell}, \citenamefont
		{Bull},\ and\ \citenamefont {Takagi}}]{19TakayamaPRB}%
	\BibitemOpen
	\bibfield  {author} {\bibinfo {author} {\bibfnamefont {T.}~\bibnamefont
			{Takayama}}, \bibinfo {author} {\bibfnamefont {A.}~\bibnamefont {Krajewska}},
		\bibinfo {author} {\bibfnamefont {A.~S.}\ \bibnamefont {Gibbs}}, \bibinfo
		{author} {\bibfnamefont {A.~N.}\ \bibnamefont {Yaresko}}, \bibinfo {author}
		{\bibfnamefont {H.}~\bibnamefont {Ishii}}, \bibinfo {author} {\bibfnamefont
			{H.}~\bibnamefont {Yamaoka}}, \bibinfo {author} {\bibfnamefont
			{K.}~\bibnamefont {Ishii}}, \bibinfo {author} {\bibfnamefont
			{N.}~\bibnamefont {Hiraoka}}, \bibinfo {author} {\bibfnamefont {N.~P.}\
			\bibnamefont {Funnell}}, \bibinfo {author} {\bibfnamefont {C.~L.}\
			\bibnamefont {Bull}},\ and\ \bibinfo {author} {\bibfnamefont
			{H.}~\bibnamefont {Takagi}},\ }\bibfield  {title} {\bibinfo {title}
		{Pressure-induced collapse of the spin-orbital {Mott} state in the
			hyperhoneycomb iridate
			$\ensuremath{\beta}\text{\ensuremath{-}}\mathrm{Li}{}_{2}\mathrm{IrO}{}_{3}$},\
	}\href {https://doi.org/10.1103/PhysRevB.99.125127} {\bibfield  {journal}
		{\bibinfo  {journal} {Phys. Rev. B}\ }\textbf {\bibinfo {volume} {99}},\
		\bibinfo {pages} {125127} (\bibinfo {year} {2019})}\BibitemShut {NoStop}%
	\bibitem [{\citenamefont {Hermann}\ \emph {et~al.}(2019)\citenamefont
		{Hermann}, \citenamefont {Ebad-Allah}, \citenamefont {Freund}, \citenamefont
		{Jesche}, \citenamefont {Tsirlin}, \citenamefont {Gegenwart},\ and\
		\citenamefont {Kuntscher}}]{19HermannPRB}%
	\BibitemOpen
	\bibfield  {author} {\bibinfo {author} {\bibfnamefont {V.}~\bibnamefont
			{Hermann}}, \bibinfo {author} {\bibfnamefont {J.}~\bibnamefont {Ebad-Allah}},
		\bibinfo {author} {\bibfnamefont {F.}~\bibnamefont {Freund}}, \bibinfo
		{author} {\bibfnamefont {A.}~\bibnamefont {Jesche}}, \bibinfo {author}
		{\bibfnamefont {A.~A.}\ \bibnamefont {Tsirlin}}, \bibinfo {author}
		{\bibfnamefont {P.}~\bibnamefont {Gegenwart}},\ and\ \bibinfo {author}
		{\bibfnamefont {C.~A.}\ \bibnamefont {Kuntscher}},\ }\bibfield  {title}
	{\bibinfo {title} {Optical signature of the pressure-induced dimerization in
			the honeycomb iridate
			$\ensuremath{\alpha}\text{\ensuremath{-}}\mathrm{Li}{}_{2}\mathrm{IrO}{}_{3}$},\
	}\href {https://doi.org/10.1103/PhysRevB.99.235116} {\bibfield  {journal}
		{\bibinfo  {journal} {Phys. Rev. B}\ }\textbf {\bibinfo {volume} {99}},\
		\bibinfo {pages} {235116} (\bibinfo {year} {2019})}\BibitemShut {NoStop}%
	\bibitem [{\citenamefont {Biffin}\ \emph {et~al.}(2014)\citenamefont {Biffin},
		\citenamefont {Johnson}, \citenamefont {Choi}, \citenamefont {Freund},
		\citenamefont {Manni}, \citenamefont {Bombardi}, \citenamefont {Manuel},
		\citenamefont {Gegenwart},\ and\ \citenamefont {Coldea}}]{14BiffinPRB}%
	\BibitemOpen
	\bibfield  {author} {\bibinfo {author} {\bibfnamefont {A.}~\bibnamefont
			{Biffin}}, \bibinfo {author} {\bibfnamefont {R.~D.}\ \bibnamefont {Johnson}},
		\bibinfo {author} {\bibfnamefont {S.}~\bibnamefont {Choi}}, \bibinfo {author}
		{\bibfnamefont {F.}~\bibnamefont {Freund}}, \bibinfo {author} {\bibfnamefont
			{S.}~\bibnamefont {Manni}}, \bibinfo {author} {\bibfnamefont
			{A.}~\bibnamefont {Bombardi}}, \bibinfo {author} {\bibfnamefont
			{P.}~\bibnamefont {Manuel}}, \bibinfo {author} {\bibfnamefont
			{P.}~\bibnamefont {Gegenwart}},\ and\ \bibinfo {author} {\bibfnamefont
			{R.}~\bibnamefont {Coldea}},\ }\bibfield  {title} {\bibinfo {title}
		{Unconventional magnetic order on the hyperhoneycomb {Kitaev} lattice in
			$\ensuremath{\beta}\text{\ensuremath{-}}\mathrm{Li}{}_{2}\mathrm{IrO}{}_{3}$:
			Full solution via magnetic resonant x-ray diffraction},\ }\href
	{https://doi.org/10.1103/PhysRevB.90.205116} {\bibfield  {journal} {\bibinfo
			{journal} {Phys. Rev. B}\ }\textbf {\bibinfo {volume} {90}},\ \bibinfo
		{pages} {205116} (\bibinfo {year} {2014})}\BibitemShut {NoStop}%
	\bibitem [{\citenamefont {Takayama}\ \emph {et~al.}(2015)\citenamefont
		{Takayama}, \citenamefont {Kato}, \citenamefont {Dinnebier}, \citenamefont
		{Nuss}, \citenamefont {Kono}, \citenamefont {Veiga}, \citenamefont {Fabbris},
		\citenamefont {Haskel},\ and\ \citenamefont {Takagi}}]{15TakayamaPRL}%
	\BibitemOpen
	\bibfield  {author} {\bibinfo {author} {\bibfnamefont {T.}~\bibnamefont
			{Takayama}}, \bibinfo {author} {\bibfnamefont {A.}~\bibnamefont {Kato}},
		\bibinfo {author} {\bibfnamefont {R.}~\bibnamefont {Dinnebier}}, \bibinfo
		{author} {\bibfnamefont {J.}~\bibnamefont {Nuss}}, \bibinfo {author}
		{\bibfnamefont {H.}~\bibnamefont {Kono}}, \bibinfo {author} {\bibfnamefont
			{L.~S.~I.}\ \bibnamefont {Veiga}}, \bibinfo {author} {\bibfnamefont
			{G.}~\bibnamefont {Fabbris}}, \bibinfo {author} {\bibfnamefont
			{D.}~\bibnamefont {Haskel}},\ and\ \bibinfo {author} {\bibfnamefont
			{H.}~\bibnamefont {Takagi}},\ }\bibfield  {title} {\bibinfo {title}
		{Hyperhoneycomb iridate
			$\ensuremath{\beta}\text{\ensuremath{-}}\mathrm{Li}{}_{2}\mathrm{IrO}{}_{3}$
			as a platform for {Kitaev} magnetism},\ }\href
	{https://doi.org/10.1103/PhysRevLett.114.077202} {\bibfield  {journal}
		{\bibinfo  {journal} {Phys. Rev. Lett.}\ }\textbf {\bibinfo {volume} {114}},\
		\bibinfo {pages} {077202} (\bibinfo {year} {2015})}\BibitemShut {NoStop}%
	\bibitem [{\citenamefont {Ruiz}\ \emph {et~al.}(2017)\citenamefont {Ruiz},
		\citenamefont {Frano}, \citenamefont {Breznay}, \citenamefont {Kimchi},
		\citenamefont {Helm}, \citenamefont {Oswald}, \citenamefont {Chan},
		\citenamefont {Birgeneau}, \citenamefont {Islam},\ and\ \citenamefont
		{Analytis}}]{17RuizNC}%
	\BibitemOpen
	\bibfield  {author} {\bibinfo {author} {\bibfnamefont {A.}~\bibnamefont
			{Ruiz}}, \bibinfo {author} {\bibfnamefont {A.}~\bibnamefont {Frano}},
		\bibinfo {author} {\bibfnamefont {N.~P.}\ \bibnamefont {Breznay}}, \bibinfo
		{author} {\bibfnamefont {I.}~\bibnamefont {Kimchi}}, \bibinfo {author}
		{\bibfnamefont {T.}~\bibnamefont {Helm}}, \bibinfo {author} {\bibfnamefont
			{I.}~\bibnamefont {Oswald}}, \bibinfo {author} {\bibfnamefont {J.~Y.}\
			\bibnamefont {Chan}}, \bibinfo {author} {\bibfnamefont {R.~J.}\ \bibnamefont
			{Birgeneau}}, \bibinfo {author} {\bibfnamefont {Z.}~\bibnamefont {Islam}},\
		and\ \bibinfo {author} {\bibfnamefont {J.~G.}\ \bibnamefont {Analytis}},\
	}\bibfield  {title} {\bibinfo {title} {Correlated states in
			$\ensuremath{\beta}\text{\ensuremath{-}}\mathrm{Li}{}_{2}\mathrm{IrO}{}_{3}$
			driven by applied magnetic fields},\ }\href
	{https://doi.org/10.1038/s41467-017-01071-9} {\bibfield  {journal} {\bibinfo
			{journal} {Nature Communications}\ }\textbf {\bibinfo {volume} {8}},\
		\bibinfo {pages} {961} (\bibinfo {year} {2017})}\BibitemShut {NoStop}%
	\bibitem [{\citenamefont {Majumder}\ \emph {et~al.}(2019)\citenamefont
		{Majumder}, \citenamefont {Freund}, \citenamefont {Dey}, \citenamefont
		{Prinz-Zwick}, \citenamefont {B\"uttgen}, \citenamefont {Skourski},
		\citenamefont {Jesche}, \citenamefont {Tsirlin},\ and\ \citenamefont
		{Gegenwart}}]{19MajumderPRM}%
	\BibitemOpen
	\bibfield  {author} {\bibinfo {author} {\bibfnamefont {M.}~\bibnamefont
			{Majumder}}, \bibinfo {author} {\bibfnamefont {F.}~\bibnamefont {Freund}},
		\bibinfo {author} {\bibfnamefont {T.}~\bibnamefont {Dey}}, \bibinfo {author}
		{\bibfnamefont {M.}~\bibnamefont {Prinz-Zwick}}, \bibinfo {author}
		{\bibfnamefont {N.}~\bibnamefont {B\"uttgen}}, \bibinfo {author}
		{\bibfnamefont {Y.}~\bibnamefont {Skourski}}, \bibinfo {author}
		{\bibfnamefont {A.}~\bibnamefont {Jesche}}, \bibinfo {author} {\bibfnamefont
			{A.~A.}\ \bibnamefont {Tsirlin}},\ and\ \bibinfo {author} {\bibfnamefont
			{P.}~\bibnamefont {Gegenwart}},\ }\bibfield  {title} {\bibinfo {title}
		{Anisotropic temperature-field phase diagram of single crystalline
			$\ensuremath{\beta}\text{\ensuremath{-}}\mathrm{Li}{}_{2}\mathrm{IrO}{}_{3}$:
			Magnetization, specific heat, and {$^{7}\mathrm{Li}$ NMR} study},\ }\href
	{https://doi.org/10.1103/PhysRevMaterials.3.074408} {\bibfield  {journal}
		{\bibinfo  {journal} {Phys. Rev. Materials}\ }\textbf {\bibinfo {volume}
			{3}},\ \bibinfo {pages} {074408} (\bibinfo {year} {2019})}\BibitemShut
	{NoStop}%
	\bibitem [{\citenamefont {Majumder}\ \emph {et~al.}(2018)\citenamefont
		{Majumder}, \citenamefont {Manna}, \citenamefont {Simutis}, \citenamefont
		{Orain}, \citenamefont {Dey}, \citenamefont {Freund}, \citenamefont {Jesche},
		\citenamefont {Khasanov}, \citenamefont {Biswas}, \citenamefont {Bykova},
		\citenamefont {Dubrovinskaia}, \citenamefont {Dubrovinsky}, \citenamefont
		{Yadav}, \citenamefont {Hozoi}, \citenamefont {Nishimoto}, \citenamefont
		{Tsirlin},\ and\ \citenamefont {Gegenwart}}]{18MajumderPRL}%
	\BibitemOpen
	\bibfield  {author} {\bibinfo {author} {\bibfnamefont {M.}~\bibnamefont
			{Majumder}}, \bibinfo {author} {\bibfnamefont {R.~S.}\ \bibnamefont {Manna}},
		\bibinfo {author} {\bibfnamefont {G.}~\bibnamefont {Simutis}}, \bibinfo
		{author} {\bibfnamefont {J.~C.}\ \bibnamefont {Orain}}, \bibinfo {author}
		{\bibfnamefont {T.}~\bibnamefont {Dey}}, \bibinfo {author} {\bibfnamefont
			{F.}~\bibnamefont {Freund}}, \bibinfo {author} {\bibfnamefont
			{A.}~\bibnamefont {Jesche}}, \bibinfo {author} {\bibfnamefont
			{R.}~\bibnamefont {Khasanov}}, \bibinfo {author} {\bibfnamefont {P.~K.}\
			\bibnamefont {Biswas}}, \bibinfo {author} {\bibfnamefont {E.}~\bibnamefont
			{Bykova}}, \bibinfo {author} {\bibfnamefont {N.}~\bibnamefont
			{Dubrovinskaia}}, \bibinfo {author} {\bibfnamefont {L.~S.}\ \bibnamefont
			{Dubrovinsky}}, \bibinfo {author} {\bibfnamefont {R.}~\bibnamefont {Yadav}},
		\bibinfo {author} {\bibfnamefont {L.}~\bibnamefont {Hozoi}}, \bibinfo
		{author} {\bibfnamefont {S.}~\bibnamefont {Nishimoto}}, \bibinfo {author}
		{\bibfnamefont {A.~A.}\ \bibnamefont {Tsirlin}},\ and\ \bibinfo {author}
		{\bibfnamefont {P.}~\bibnamefont {Gegenwart}},\ }\bibfield  {title} {\bibinfo
		{title} {Breakdown of magnetic order in the pressurized {Kitaev} iridate
			$\ensuremath{\beta}\text{\ensuremath{-}}\mathrm{Li}{}_{2}\mathrm{IrO}{}_{3}$},\
	}\href {https://doi.org/10.1103/PhysRevLett.120.237202} {\bibfield  {journal}
		{\bibinfo  {journal} {Phys. Rev. Lett.}\ }\textbf {\bibinfo {volume} {120}},\
		\bibinfo {pages} {237202} (\bibinfo {year} {2018})}\BibitemShut {NoStop}%
	\bibitem [{\citenamefont {Choi}\ \emph {et~al.}(2020)\citenamefont {Choi},
		\citenamefont {Kim}, \citenamefont {Kim}, \citenamefont {Krajewska},
		\citenamefont {Kim}, \citenamefont {Minola}, \citenamefont {Takayama},
		\citenamefont {Takagi}, \citenamefont {Haule}, \citenamefont {Vanderbilt},\
		and\ \citenamefont {Keimer}}]{choi2020}%
	\BibitemOpen
	\bibfield  {author} {\bibinfo {author} {\bibfnamefont {S.}~\bibnamefont
			{Choi}}, \bibinfo {author} {\bibfnamefont {H.-S.}\ \bibnamefont {Kim}},
		\bibinfo {author} {\bibfnamefont {H.-H.}\ \bibnamefont {Kim}}, \bibinfo
		{author} {\bibfnamefont {A.}~\bibnamefont {Krajewska}}, \bibinfo {author}
		{\bibfnamefont {G.}~\bibnamefont {Kim}}, \bibinfo {author} {\bibfnamefont
			{M.}~\bibnamefont {Minola}}, \bibinfo {author} {\bibfnamefont
			{T.}~\bibnamefont {Takayama}}, \bibinfo {author} {\bibfnamefont
			{H.}~\bibnamefont {Takagi}}, \bibinfo {author} {\bibfnamefont
			{K.}~\bibnamefont {Haule}}, \bibinfo {author} {\bibfnamefont
			{D.}~\bibnamefont {Vanderbilt}},\ and\ \bibinfo {author} {\bibfnamefont
			{B.}~\bibnamefont {Keimer}},\ }\bibfield  {title} {\bibinfo {title} {Lattice
			dynamics and structural transition of the hyperhoneycomb iridate
			{$\beta$-Li$_2$IrO$_3$} investigated by high-pressure {Raman} scattering},\
	}\href {https://doi.org/10.1103/PhysRevB.101.054102} {\bibfield  {journal}
		{\bibinfo  {journal} {Phys. Rev. B}\ }\textbf {\bibinfo {volume} {101}},\
		\bibinfo {pages} {054102} (\bibinfo {year} {2020})}\BibitemShut {NoStop}%
	\bibitem [{\citenamefont {Veiga}\ \emph {et~al.}(2019)\citenamefont {Veiga},
		\citenamefont {Glazyrin}, \citenamefont {Fabbris}, \citenamefont {Dashwood},
		\citenamefont {Vale}, \citenamefont {Park}, \citenamefont {Etter},
		\citenamefont {Irifune}, \citenamefont {Pascarelli}, \citenamefont
		{McMorrow}, \citenamefont {Takayama}, \citenamefont {Takagi},\ and\
		\citenamefont {Haskel}}]{19VeigaPRB}%
	\BibitemOpen
	\bibfield  {author} {\bibinfo {author} {\bibfnamefont {L.~S.~I.}\
			\bibnamefont {Veiga}}, \bibinfo {author} {\bibfnamefont {K.}~\bibnamefont
			{Glazyrin}}, \bibinfo {author} {\bibfnamefont {G.}~\bibnamefont {Fabbris}},
		\bibinfo {author} {\bibfnamefont {C.~D.}\ \bibnamefont {Dashwood}}, \bibinfo
		{author} {\bibfnamefont {J.~G.}\ \bibnamefont {Vale}}, \bibinfo {author}
		{\bibfnamefont {H.}~\bibnamefont {Park}}, \bibinfo {author} {\bibfnamefont
			{M.}~\bibnamefont {Etter}}, \bibinfo {author} {\bibfnamefont
			{T.}~\bibnamefont {Irifune}}, \bibinfo {author} {\bibfnamefont
			{S.}~\bibnamefont {Pascarelli}}, \bibinfo {author} {\bibfnamefont {D.~F.}\
			\bibnamefont {McMorrow}}, \bibinfo {author} {\bibfnamefont {T.}~\bibnamefont
			{Takayama}}, \bibinfo {author} {\bibfnamefont {H.}~\bibnamefont {Takagi}},\
		and\ \bibinfo {author} {\bibfnamefont {D.}~\bibnamefont {Haskel}},\
	}\bibfield  {title} {\bibinfo {title} {Pressure-induced structural
			dimerization in the hyperhoneycomb iridate
			$\ensuremath{\beta}\text{\ensuremath{-}}\mathrm{Li}{}_{2}\mathrm{IrO}{}_{3}$
			at low temperatures},\ }\href {https://doi.org/10.1103/PhysRevB.100.064104}
	{\bibfield  {journal} {\bibinfo  {journal} {Phys. Rev. B}\ }\textbf {\bibinfo
			{volume} {100}},\ \bibinfo {pages} {064104} (\bibinfo {year}
		{2019})}\BibitemShut {NoStop}%
	\bibitem [{\citenamefont {Seidler}()}]{Max}%
	\BibitemOpen
	\bibfield  {author} {\bibinfo {author} {\bibfnamefont {M.}~\bibnamefont
			{Seidler}},\ }\href {https://github.com/miile7/mpms-analyzer} {\bibinfo
		{title} {{MPMS Analyzer software}}}\BibitemShut {NoStop}%
	\bibitem [{\citenamefont {Tsirlin}\ \emph {et~al.}(2021)\citenamefont
		{Tsirlin}, \citenamefont {Zubtsovskii},\ and\ \citenamefont {Uykur}}]{esrf}%
	\BibitemOpen
	\bibfield  {author} {\bibinfo {author} {\bibfnamefont {A.~A.}\ \bibnamefont
			{Tsirlin}}, \bibinfo {author} {\bibfnamefont {A.~O.}\ \bibnamefont
			{Zubtsovskii}},\ and\ \bibinfo {author} {\bibfnamefont {E.}~\bibnamefont
			{Uykur}},\ }\bibfield  {title} {\bibinfo {title} {High-resolution x-ray
			diffraction on decompressed samples of {$\beta$-Li$_2$IrO$_3$}},\ }\bibfield
	{journal} {\bibinfo  {journal} {European Synchrotron Radiation Facility
			(ESRF)}\ }\href {https://doi.org/10.15151/ESRF-DC-568780168}
	{10.15151/ESRF-DC-568780168} (\bibinfo {year} {2021})\BibitemShut {NoStop}%
	\bibitem [{\citenamefont {Pet{\u r}{\'\i}{\u c}ek}\ \emph
		{et~al.}(2014)\citenamefont {Pet{\u r}{\'\i}{\u c}ek}, \citenamefont {Du{\u
				s}ek},\ and\ \citenamefont {Palatinus}}]{jana2006}%
	\BibitemOpen
	\bibfield  {author} {\bibinfo {author} {\bibfnamefont {V.}~\bibnamefont
			{Pet{\u r}{\'\i}{\u c}ek}}, \bibinfo {author} {\bibfnamefont
			{M.}~\bibnamefont {Du{\u s}ek}},\ and\ \bibinfo {author} {\bibfnamefont
			{L.}~\bibnamefont {Palatinus}},\ }\bibfield  {title} {\bibinfo {title}
		{Crystallographic computing system {JANA2006}: General features},\ }\href
	{https://doi.org/10.1515/zkri-2014-1737} {\bibfield  {journal} {\bibinfo
			{journal} {Z. Krist.}\ }\textbf {\bibinfo {volume} {229}},\ \bibinfo {pages}
		{345} (\bibinfo {year} {2014})}\BibitemShut {NoStop}%
	\bibitem [{\citenamefont {Kresse}\ and\ \citenamefont
		{Furthm\"uller}(1996{\natexlab{a}})}]{vasp1}%
	\BibitemOpen
	\bibfield  {author} {\bibinfo {author} {\bibfnamefont {G.}~\bibnamefont
			{Kresse}}\ and\ \bibinfo {author} {\bibfnamefont {J.}~\bibnamefont
			{Furthm\"uller}},\ }\bibfield  {title} {\bibinfo {title} {Efficiency of
			\textit{ab-initio} total energy calculations for metals and semiconductors
			using a plane-wave basis set},\ }\href
	{https://doi.org/http://dx.doi.org/10.1016/0927-0256(96)00008-0} {\bibfield
		{journal} {\bibinfo  {journal} {Comput. Mater. Sci.}\ }\textbf {\bibinfo
			{volume} {6}},\ \bibinfo {pages} {15} (\bibinfo {year}
		{1996}{\natexlab{a}})}\BibitemShut {NoStop}%
	\bibitem [{\citenamefont {Kresse}\ and\ \citenamefont
		{Furthm\"uller}(1996{\natexlab{b}})}]{vasp2}%
	\BibitemOpen
	\bibfield  {author} {\bibinfo {author} {\bibfnamefont {G.}~\bibnamefont
			{Kresse}}\ and\ \bibinfo {author} {\bibfnamefont {J.}~\bibnamefont
			{Furthm\"uller}},\ }\bibfield  {title} {\bibinfo {title} {Efficient iterative
			schemes for \textit{ab initio} total-energy calculations using a plane-wave
			basis set},\ }\href {https://doi.org/10.1103/PhysRevB.54.11169} {\bibfield
		{journal} {\bibinfo  {journal} {Phys. Rev. B}\ }\textbf {\bibinfo {volume}
			{54}},\ \bibinfo {pages} {11169} (\bibinfo {year}
		{1996}{\natexlab{b}})}\BibitemShut {NoStop}%
	\bibitem [{\citenamefont {Perdew}\ \emph {et~al.}(2008)\citenamefont {Perdew},
		\citenamefont {Ruzsinszky}, \citenamefont {Csonka}, \citenamefont {Vydrov},
		\citenamefont {Scuseria}, \citenamefont {Constantin}, \citenamefont {Zhou},\
		and\ \citenamefont {Burke}}]{pbesol}%
	\BibitemOpen
	\bibfield  {author} {\bibinfo {author} {\bibfnamefont {J.~P.}\ \bibnamefont
			{Perdew}}, \bibinfo {author} {\bibfnamefont {A.}~\bibnamefont {Ruzsinszky}},
		\bibinfo {author} {\bibfnamefont {G.~I.}\ \bibnamefont {Csonka}}, \bibinfo
		{author} {\bibfnamefont {O.~A.}\ \bibnamefont {Vydrov}}, \bibinfo {author}
		{\bibfnamefont {G.~E.}\ \bibnamefont {Scuseria}}, \bibinfo {author}
		{\bibfnamefont {L.~A.}\ \bibnamefont {Constantin}}, \bibinfo {author}
		{\bibfnamefont {X.}~\bibnamefont {Zhou}},\ and\ \bibinfo {author}
		{\bibfnamefont {K.}~\bibnamefont {Burke}},\ }\bibfield  {title} {\bibinfo
		{title} {Restoring the density-gradient expansion for exchange in solids and
			surfaces},\ }\href {https://doi.org/10.1103/PhysRevLett.100.136406}
	{\bibfield  {journal} {\bibinfo  {journal} {Phys. Rev. Lett.}\ }\textbf
		{\bibinfo {volume} {100}},\ \bibinfo {pages} {136406} (\bibinfo {year}
		{2008})}\BibitemShut {NoStop}%
	\bibitem [{\citenamefont {Koepernik}\ and\ \citenamefont
		{Eschrig}(1999)}]{fplo}%
	\BibitemOpen
	\bibfield  {author} {\bibinfo {author} {\bibfnamefont {K.}~\bibnamefont
			{Koepernik}}\ and\ \bibinfo {author} {\bibfnamefont {H.}~\bibnamefont
			{Eschrig}},\ }\bibfield  {title} {\bibinfo {title} {Full-potential
			nonorthogonal local-orbital minimum-basis band-structure scheme},\ }\href
	{https://doi.org/10.1103/PhysRevB.59.1743} {\bibfield  {journal} {\bibinfo
			{journal} {Phys. Rev. B}\ }\textbf {\bibinfo {volume} {59}},\ \bibinfo
		{pages} {1743} (\bibinfo {year} {1999})}\BibitemShut {NoStop}%
	\bibitem [{\citenamefont {Veiga}\ \emph {et~al.}(2017)\citenamefont {Veiga},
		\citenamefont {Etter}, \citenamefont {Glazyrin}, \citenamefont {Sun},
		\citenamefont {Escanhoela}, \citenamefont {Fabbris}, \citenamefont
		{Mardegan}, \citenamefont {Malavi}, \citenamefont {Deng}, \citenamefont
		{Stavropoulos}, \citenamefont {Kee}, \citenamefont {Yang}, \citenamefont {van
			Veenendaal}, \citenamefont {Schilling}, \citenamefont {Takayama},
		\citenamefont {Takagi},\ and\ \citenamefont {Haskel}}]{17VeigaPRB}%
	\BibitemOpen
	\bibfield  {author} {\bibinfo {author} {\bibfnamefont {L.~S.~I.}\
			\bibnamefont {Veiga}}, \bibinfo {author} {\bibfnamefont {M.}~\bibnamefont
			{Etter}}, \bibinfo {author} {\bibfnamefont {K.}~\bibnamefont {Glazyrin}},
		\bibinfo {author} {\bibfnamefont {F.}~\bibnamefont {Sun}}, \bibinfo {author}
		{\bibfnamefont {C.~A.}\ \bibnamefont {Escanhoela}}, \bibinfo {author}
		{\bibfnamefont {G.}~\bibnamefont {Fabbris}}, \bibinfo {author} {\bibfnamefont
			{J.~R.~L.}\ \bibnamefont {Mardegan}}, \bibinfo {author} {\bibfnamefont
			{P.~S.}\ \bibnamefont {Malavi}}, \bibinfo {author} {\bibfnamefont
			{Y.}~\bibnamefont {Deng}}, \bibinfo {author} {\bibfnamefont {P.~P.}\
			\bibnamefont {Stavropoulos}}, \bibinfo {author} {\bibfnamefont {H.-Y.}\
			\bibnamefont {Kee}}, \bibinfo {author} {\bibfnamefont {W.~G.}\ \bibnamefont
			{Yang}}, \bibinfo {author} {\bibfnamefont {M.}~\bibnamefont {van
				Veenendaal}}, \bibinfo {author} {\bibfnamefont {J.~S.}\ \bibnamefont
			{Schilling}}, \bibinfo {author} {\bibfnamefont {T.}~\bibnamefont {Takayama}},
		\bibinfo {author} {\bibfnamefont {H.}~\bibnamefont {Takagi}},\ and\ \bibinfo
		{author} {\bibfnamefont {D.}~\bibnamefont {Haskel}},\ }\bibfield  {title}
	{\bibinfo {title} {Pressure tuning of bond-directional exchange interactions
			and magnetic frustration in the hyperhoneycomb iridate
			$\ensuremath{\beta}\text{\ensuremath{-}}\mathrm{Li}{}_{2}\mathrm{IrO}{}_{3}$},\
	}\href {https://doi.org/10.1103/PhysRevB.96.140402} {\bibfield  {journal}
		{\bibinfo  {journal} {Phys. Rev. B}\ }\textbf {\bibinfo {volume} {96}},\
		\bibinfo {pages} {140402} (\bibinfo {year} {2017})}\BibitemShut {NoStop}%
	\bibitem [{\citenamefont {Andreica}\ \emph {et~al.}(2000)\citenamefont
		{Andreica}, \citenamefont {Cavadini}, \citenamefont {G\"udel}, \citenamefont
		{Gygax}, \citenamefont {Kr\"amer}, \citenamefont {Pinkpank},\ and\
		\citenamefont {Schenck}}]{00AndreicaPB}%
	\BibitemOpen
	\bibfield  {author} {\bibinfo {author} {\bibfnamefont {D.}~\bibnamefont
			{Andreica}}, \bibinfo {author} {\bibfnamefont {N.}~\bibnamefont {Cavadini}},
		\bibinfo {author} {\bibfnamefont {H.~U.}\ \bibnamefont {G\"udel}}, \bibinfo
		{author} {\bibfnamefont {F.~N.}\ \bibnamefont {Gygax}}, \bibinfo {author}
		{\bibfnamefont {K.}~\bibnamefont {Kr\"amer}}, \bibinfo {author}
		{\bibfnamefont {M.}~\bibnamefont {Pinkpank}},\ and\ \bibinfo {author}
		{\bibfnamefont {A.}~\bibnamefont {Schenck}},\ }\bibfield  {title} {\bibinfo
		{title} {Muon-induced break up of spin-singlet pairs in the double-chain
			compound $\ensuremath\mathrm{KCu}\mathrm{Cl}{}_{3}$},\ }\href
	{https://doi.org/10.1016/S0921-4526(00)00355-0} {\bibfield  {journal}
		{\bibinfo  {journal} {Physica B}\ }\textbf {\bibinfo {volume} {289}},\
		\bibinfo {pages} {176} (\bibinfo {year} {2000})}\BibitemShut {NoStop}%
	\bibitem [{\citenamefont {Cavadini}\ \emph {et~al.}(2003)\citenamefont
		{Cavadini}, \citenamefont {Andreica}, \citenamefont {Gygax}, \citenamefont
		{Schenck}, \citenamefont {Kr\"amer}, \citenamefont {G\"udel}, \citenamefont
		{Mutka},\ and\ \citenamefont {Wildes}}]{03CavadiniPB}%
	\BibitemOpen
	\bibfield  {author} {\bibinfo {author} {\bibfnamefont {N.}~\bibnamefont
			{Cavadini}}, \bibinfo {author} {\bibfnamefont {D.}~\bibnamefont {Andreica}},
		\bibinfo {author} {\bibfnamefont {F.~N.}\ \bibnamefont {Gygax}}, \bibinfo
		{author} {\bibfnamefont {A.}~\bibnamefont {Schenck}}, \bibinfo {author}
		{\bibfnamefont {K.}~\bibnamefont {Kr\"amer}}, \bibinfo {author}
		{\bibfnamefont {H.-U.}\ \bibnamefont {G\"udel}}, \bibinfo {author}
		{\bibfnamefont {H.}~\bibnamefont {Mutka}},\ and\ \bibinfo {author}
		{\bibfnamefont {A.}~\bibnamefont {Wildes}},\ }\bibfield  {title} {\bibinfo
		{title} {Local spin susceptibility in
			$\ensuremath\mathrm{KCu}\mathrm{Cl}{}_{3}$},\ }\href
	{https://doi.org/10.1016/S0921-4526(03)00186-8} {\bibfield  {journal}
		{\bibinfo  {journal} {Physica B}\ }\textbf {\bibinfo {volume} {335}},\
		\bibinfo {pages} {37} (\bibinfo {year} {2003})}\BibitemShut {NoStop}%
	\bibitem [{\citenamefont {Lappas}\ \emph {et~al.}(2003)\citenamefont {Lappas},
		\citenamefont {Schenck},\ and\ \citenamefont {Prassides}}]{03CavadiniPB2}%
	\BibitemOpen
	\bibfield  {author} {\bibinfo {author} {\bibfnamefont {A.}~\bibnamefont
			{Lappas}}, \bibinfo {author} {\bibfnamefont {A.}~\bibnamefont {Schenck}},\
		and\ \bibinfo {author} {\bibfnamefont {K.}~\bibnamefont {Prassides}},\
	}\bibfield  {title} {\bibinfo {title} {Spin-freezing in the two-dimensional
			spin-gap systems
			$\ensuremath\mathrm{Sr}\mathrm{Cu}{}_{2-x}\mathrm{Mg}{}_{x}\mathrm{(BO{}_{3})}{}_{2}$
			(\ensuremath{x}=0,0.04,0.12)},\ }\href
	{https://doi.org/10.1016/S0921-4526(02)01669-1} {\bibfield  {journal}
		{\bibinfo  {journal} {Physica B}\ }\textbf {\bibinfo {volume} {326}},\
		\bibinfo {pages} {431} (\bibinfo {year} {2003})}\BibitemShut {NoStop}%
	\bibitem [{\citenamefont {Ruiz}\ \emph {et~al.}(2020)\citenamefont {Ruiz},
		\citenamefont {Nagarajan}, \citenamefont {Vranas}, \citenamefont {Lopez},
		\citenamefont {McCandless}, \citenamefont {Kimchi}, \citenamefont {Chan},
		\citenamefont {Breznay}, \citenamefont {Fra\~n\'o}, \citenamefont
		{Frandsen},\ and\ \citenamefont {Analytis}}]{ruiz2020}%
	\BibitemOpen
	\bibfield  {author} {\bibinfo {author} {\bibfnamefont {A.}~\bibnamefont
			{Ruiz}}, \bibinfo {author} {\bibfnamefont {V.}~\bibnamefont {Nagarajan}},
		\bibinfo {author} {\bibfnamefont {M.}~\bibnamefont {Vranas}}, \bibinfo
		{author} {\bibfnamefont {G.}~\bibnamefont {Lopez}}, \bibinfo {author}
		{\bibfnamefont {G.~T.}\ \bibnamefont {McCandless}}, \bibinfo {author}
		{\bibfnamefont {I.}~\bibnamefont {Kimchi}}, \bibinfo {author} {\bibfnamefont
			{J.~Y.}\ \bibnamefont {Chan}}, \bibinfo {author} {\bibfnamefont {N.~P.}\
			\bibnamefont {Breznay}}, \bibinfo {author} {\bibfnamefont {A.}~\bibnamefont
			{Fra\~n\'o}}, \bibinfo {author} {\bibfnamefont {B.~A.}\ \bibnamefont
			{Frandsen}},\ and\ \bibinfo {author} {\bibfnamefont {J.~G.}\ \bibnamefont
			{Analytis}},\ }\bibfield  {title} {\bibinfo {title} {High-temperature
			magnetic anomaly in the {Kitaev} hyperhoneycomb compound
			{$\beta$-Li$_2$IrO$_3$}},\ }\href
	{https://doi.org/10.1103/PhysRevB.101.075112} {\bibfield  {journal} {\bibinfo
			{journal} {Phys. Rev. B}\ }\textbf {\bibinfo {volume} {101}},\ \bibinfo
		{pages} {075112} (\bibinfo {year} {2020})}\BibitemShut {NoStop}%
	\bibitem [{\citenamefont {Antonov}\ \emph {et~al.}(2018)\citenamefont
		{Antonov}, \citenamefont {Uba},\ and\ \citenamefont {Uba}}]{antonov2018}%
	\BibitemOpen
	\bibfield  {author} {\bibinfo {author} {\bibfnamefont {V.~N.}\ \bibnamefont
			{Antonov}}, \bibinfo {author} {\bibfnamefont {S.}~\bibnamefont {Uba}},\ and\
		\bibinfo {author} {\bibfnamefont {L.}~\bibnamefont {Uba}},\ }\bibfield
	{title} {\bibinfo {title} {Electronic structure and x-ray magnetic circular
			dichroism in the hyperhoneycomb iridate {$\beta$-Li$_2$IrO$_3$}},\ }\href
	{https://doi.org/10.1103/PhysRevB.98.245113} {\bibfield  {journal} {\bibinfo
			{journal} {Phys. Rev. B}\ }\textbf {\bibinfo {volume} {98}},\ \bibinfo
		{pages} {245113} (\bibinfo {year} {2018})}\BibitemShut {NoStop}%
	\bibitem [{\citenamefont {Antonov}\ \emph {et~al.}(2021)\citenamefont
		{Antonov}, \citenamefont {Kukusta}, \citenamefont {Uba}, \citenamefont
		{Bonda},\ and\ \citenamefont {Uba}}]{antonov2021}%
	\BibitemOpen
	\bibfield  {author} {\bibinfo {author} {\bibfnamefont {V.~N.}\ \bibnamefont
			{Antonov}}, \bibinfo {author} {\bibfnamefont {D.~A.}\ \bibnamefont
			{Kukusta}}, \bibinfo {author} {\bibfnamefont {L.}~\bibnamefont {Uba}},
		\bibinfo {author} {\bibfnamefont {A.}~\bibnamefont {Bonda}},\ and\ \bibinfo
		{author} {\bibfnamefont {S.}~\bibnamefont {Uba}},\ }\bibfield  {title}
	{\bibinfo {title} {Resonant inelastic x-ray scattering spectra in the
			hyperhoneycomb iridate {$\beta$-Li$_2$IrO$_3$}: First-principles
			calculations},\ }\href {https://doi.org/10.1103/PhysRevB.103.235127}
	{\bibfield  {journal} {\bibinfo  {journal} {Phys. Rev. B}\ }\textbf {\bibinfo
			{volume} {103}},\ \bibinfo {pages} {235127} (\bibinfo {year}
		{2021})}\BibitemShut {NoStop}%
	\bibitem [{\citenamefont {Rau}\ \emph {et~al.}(2014)\citenamefont {Rau},
		\citenamefont {Lee},\ and\ \citenamefont {Kee}}]{rau2014}%
	\BibitemOpen
	\bibfield  {author} {\bibinfo {author} {\bibfnamefont {J.~G.}\ \bibnamefont
			{Rau}}, \bibinfo {author} {\bibfnamefont {E.~K.-H.}\ \bibnamefont {Lee}},\
		and\ \bibinfo {author} {\bibfnamefont {H.-Y.}\ \bibnamefont {Kee}},\
	}\bibfield  {title} {\bibinfo {title} {Generic spin model for the honeycomb
			iridates beyond the {Kitaev} limit},\ }\href
	{https://doi.org/10.1103/PhysRevLett.112.077204} {\bibfield  {journal}
		{\bibinfo  {journal} {Phys. Rev. Lett.}\ }\textbf {\bibinfo {volume} {112}},\
		\bibinfo {pages} {077204} (\bibinfo {year} {2014})}\BibitemShut {NoStop}%
	\bibitem [{\citenamefont {Winter}\ \emph {et~al.}(2016)\citenamefont {Winter},
		\citenamefont {Li}, \citenamefont {Jeschke},\ and\ \citenamefont
		{Valent{\'\i}}}]{winter2016}%
	\BibitemOpen
	\bibfield  {author} {\bibinfo {author} {\bibfnamefont {S.}~\bibnamefont
			{Winter}}, \bibinfo {author} {\bibfnamefont {Y.}~\bibnamefont {Li}}, \bibinfo
		{author} {\bibfnamefont {H.}~\bibnamefont {Jeschke}},\ and\ \bibinfo {author}
		{\bibfnamefont {R.}~\bibnamefont {Valent{\'\i}}},\ }\bibfield  {title}
	{\bibinfo {title} {Challenges in design of {Kitaev} materials: Magnetic
			interactions from competing energy scales},\ }\href
	{https://doi.org/10.1103/PhysRevB.93.214431} {\bibfield  {journal} {\bibinfo
			{journal} {Phys. Rev. B}\ }\textbf {\bibinfo {volume} {93}},\ \bibinfo
		{pages} {214431} (\bibinfo {year} {2016})}\BibitemShut {NoStop}%
	\bibitem [{\citenamefont {Majumder}\ \emph {et~al.}(2020)\citenamefont
		{Majumder}, \citenamefont {Prinz-Zwick}, \citenamefont {Reschke},
		\citenamefont {Zubtsovskii}, \citenamefont {Dey}, \citenamefont {Freund},
		\citenamefont {B\"uttgen}, \citenamefont {Jesche}, \citenamefont
		{K\'ezsm\'arki}, \citenamefont {Tsirlin},\ and\ \citenamefont
		{Gegenwart}}]{20MajumderPRB}%
	\BibitemOpen
	\bibfield  {author} {\bibinfo {author} {\bibfnamefont {M.}~\bibnamefont
			{Majumder}}, \bibinfo {author} {\bibfnamefont {M.}~\bibnamefont
			{Prinz-Zwick}}, \bibinfo {author} {\bibfnamefont {S.}~\bibnamefont
			{Reschke}}, \bibinfo {author} {\bibfnamefont {A.}~\bibnamefont
			{Zubtsovskii}}, \bibinfo {author} {\bibfnamefont {T.}~\bibnamefont {Dey}},
		\bibinfo {author} {\bibfnamefont {F.}~\bibnamefont {Freund}}, \bibinfo
		{author} {\bibfnamefont {N.}~\bibnamefont {B\"uttgen}}, \bibinfo {author}
		{\bibfnamefont {A.}~\bibnamefont {Jesche}}, \bibinfo {author} {\bibfnamefont
			{I.}~\bibnamefont {K\'ezsm\'arki}}, \bibinfo {author} {\bibfnamefont {A.~A.}\
			\bibnamefont {Tsirlin}},\ and\ \bibinfo {author} {\bibfnamefont
			{P.}~\bibnamefont {Gegenwart}},\ }\bibfield  {title} {\bibinfo {title} {Field
			evolution of low-energy excitations in the hyperhoneycomb magnet
			$\ensuremath{\beta}\text{\ensuremath{-}}\mathrm{Li}{}_{2}\mathrm{IrO}{}_{3}$},\
	}\href {https://doi.org/10.1103/PhysRevB.101.214417} {\bibfield  {journal}
		{\bibinfo  {journal} {Phys. Rev. B}\ }\textbf {\bibinfo {volume} {101}},\
		\bibinfo {pages} {214417} (\bibinfo {year} {2020})}\BibitemShut {NoStop}%
	\bibitem [{\citenamefont {Rousochatzakis}\ and\ \citenamefont
		{Perkins}(2018)}]{rousochatzakis2018}%
	\BibitemOpen
	\bibfield  {author} {\bibinfo {author} {\bibfnamefont {I.}~\bibnamefont
			{Rousochatzakis}}\ and\ \bibinfo {author} {\bibfnamefont {N.~B.}\
			\bibnamefont {Perkins}},\ }\bibfield  {title} {\bibinfo {title} {Magnetic
			field induced evolution of intertwined orders in the {Kitaev} magnet
			{$\beta$-Li$_2$IrO$_3$}},\ }\href
	{https://doi.org/10.1103/PhysRevB.97.174423} {\bibfield  {journal} {\bibinfo
			{journal} {Phys. Rev. B}\ }\textbf {\bibinfo {volume} {97}},\ \bibinfo
		{pages} {174423} (\bibinfo {year} {2018})}\BibitemShut {NoStop}%
\end{thebibliography}
\end{document}